%% file: main.tex
\documentclass[manuscript, screen]{acmart}

\usepackage{./colors}
\usepackage{./packages}

\input{./parts/setup}
\input{./parts/abstract}

\begin{document}

\maketitle

\input{./parts/introduction}
\input{./parts/background}

\input{./parts/motivations}

\input{./parts/architecture}

\input{./parts/comparison}
\input{./parts/mappings}
\input{./parts/results}
\input{./parts/conclusions}

\section*{Acknowledgements}
This work has been supported by IMEC, Leuven, Belgium.

\bibliographystyle{ACM-Reference-Format}
\bibliography{references.bib}

\end{document}

%% file: parts/setup.tex
\setcopyright{acmlicensed}
\copyrightyear{2025}
\acmYear{2025}
\acmDOI{XXXXXXX.XXXXXXX}

\title{Addressing memory bandwidth scalability in vector processors for streaming applications}

\author{Jordi Altay\'o}
\email{jordiag@kth.se}
\orcid{0000-0002-7693-6994}
\author{Paul Delestrac}
\email{pauldel@kth.se}
\orcid{0000-0002-7476-1422}
\author{Ahmed Hemani}
\email{hemani@kth.se}
\orcid{0000-0003-0565-9376}
\affiliation{%
  \institution{KTH - Royal Institute of Technology}
  \city{Stockholm}
  \country{Sweden}
}

\author{David Novo}
\email{david.novo@cnrs.fr}
\orcid{0000-0002-5510-4152}
\affiliation{%
  \institution{LIRMM - Laboratoire d'informatique, de robotique et de microélectronique de Montpellier}
  \city{Montpellier}
  \country{France}}

\author{Simey Yang}
\authornote{The work was done while the author was affiliated with IMEC.}
\email{simey.yang@imec.be}
\orcid{0000-0002-0130-8176}
\affiliation{%
  \institution{South China University of Technology}
  \city{Guangzhou}
  \country{China}}
\author{Debjyoti Bhattacharjee}
\email{debjoti.bhattacharjee@imec.be}
\orcid{0000-0001-6561-8934}
\author{Francky Cathoor}
\email{francky.cathoor@imec.be}
\orcid{0000-0002-3599-8515}
\affiliation{%
  \institution{IMEC - Interuniversity Microelectronics Centre}
  \city{Louven}
  \country{Belgium}}

\ccsdesc[500]{Hardware~Application specific instruction set processors}

\keywords{ASIP, DSIP, Machine Learning, Accelerator, ASIC}

%% file: parts/abstract.tex
\begin{abstract}
As the size of artificial intelligence and machine learning (AI/ML) models and datasets grows, the memory bandwidth becomes a
critical bottleneck. The paper presents a novel extended memory hierarchy that addresses some major memory bandwidth challenges
in data-parallel AI/ML applications.
While data-parallel architectures like GPUs and neural network accelerators have improved power performance compared to
traditional CPUs, they can still be significantly bottlenecked by their memory bandwidth, especially when the data reuse in the loop kernels is limited.
Systolic arrays (SAs) and GPUs attempt to mitigate the memory bandwidth bottleneck but can still become memory bandwidth
throttled when the amount of data reuse is not sufficient to confine data access mostly to the local memories near to the processing.
To mitigate this, the proposed architecture introduces three levels of on-chip memory --- local, intermediate, and global --- with an
ultra-wide register and data-shufflers to improve versatility and adaptivity to varying data-parallel applications.
The paper explains the innovations at a conceptual level and presents a detailed description of the architecture innovations.
We also map a representative data-parallel application, like a convolutional neural network (CNN), to the proposed
architecture and quantify the benefits vis-a-vis GPUs and repersentative accelerators based on systolic arrays and
vector processors.
\end{abstract}

%% file: parts/introduction.tex
\section{Introduction\label{sec:introduction}}
The \emph{von Neumann} bottleneck~\cite{Backus2007} has been a long-standing challenge in computer architecture. The
separation between memory and computation has lead to the current memory-dominated architectures, where data movement
has become the main bottleneck \cite{Kanev2015,boroumand2021google}, both for performance and energy consumption. This
effect is significantly pronounced in data-parallel applications, such as artificial intelligence and machine learning
(AI/ML), where the size of the models and datasets used in these applications has been growing
exponentially~\cite{Villalobos2022, boroumand2021google, boroumand2021mitigating, yu2021compute, kwon2018beyond}. When
memory bandwidth becomes the bottleneck, the performance gains achieved by exploiting the data-level parallelism in
these applications are overshadowed by memory accesses, leading to stalled processing elements and wasted
energy~\cite{chen2014diannao, 10.1145/3007787.3001177}.


An ample number of architectures have been proposed as potential solutions to the memory bandwidth bottleneck. Among
these, systolic arrays (SAs), single-instruction multiple-data (SIMD), single-instruction multiple-thread (SIMT), and
vector processors have been the most prominent. Different flavours of them have shown potential strengths in these
applications. Eyeriss \cite{10.1109/ISCA.2016.40} and TPU \cite{10.1145/3173162.3173177} are representative examples of
SA-based, while ARA \cite{Perotti2022} and Nvidia's Ampere GPU architecture~\cite{ampere2020} represent vector and SIMT
architectures, respectively.

We show how heavy reliance on data reuse can lead to memory throttling when the available level of reuse becomes
insufficient to feed the memory hierarchy and to keep the data access mostly to the local memories near the processing
\cite{delestrac2024multi, Dadu2019}. We also show how the scalability of these architectures is limited by different
factors, such as the 2D organization of the PEs in SAs, or the lack of flexibility in the interconnect of the current
vector processors.

We propose a vector architecture extension that addresses these limitations by introducing a novel extended memory
hierarchy with scalable bandwidth, and does not rely mainly on data reuse for achieving high performance. So, it also
performs well when data reuse factors are quite low, albeit at the expense of initiating more costly data move
operations in that case. The architecture is based on a 1D organization of the PEs but introduces extra elements in the
memory hierarchy that allow it to maintain the scalability of vector processors while maintaining the efficiency on data
movement of systolic arrays. In particular, it introduces an ultra-wide single-port register and a specialized wide data
shuffler.

\subsection{Scope}
This work focuses on the architectural aspects and how the memory bandwidth limitations are addressed. We do not delve
into the details of the implementation of the processing elements beyond the details required to understand the data
movement within them. Furthermore, to maintain the length of the paper within the limits and to conceptually split the
work between bandwidth and energy efficiency, we do not perform a detailed analysis and comparison of the energy
efficiency. However, we still provide a qualitative explanation on why the proposed memory hierarchy is, at least, as energy
efficient as an equivalently sized alternative, with potential for obtaining significant gains when scaling the
architecture. We leave a detailed analysis of the energy efficiency for future work. 

\subsection{Contributions and paper organisation}
The contributions of this work are organized as follows:

\begin{itemize}
  \item We identify the limitations of the current architectures based on GPUs, systolic arrays and vector processors, and
    motivate the need for the addition of the architectural elements proposed in this work in Section~\ref{sec:motivations}.

  \item We propose an architecture extension that addresses these limitations by introducing a modified memory hierarchy
    that is scalable and does not rely mainly on data reuse for achieving high performance. So, it even functions under
    low data reuse factors. We also include the processing elements emphasizing the data shufflers in
    Section~\ref{sec:detailed_description}.
  
  \item We present a detailed comparison at the architectural level with the four architectures mentioned above: Eyeriss,
    TPU, ARA, and GPUs in Section~\ref{sec:comparison}.
  
  \item We show how a representative data-parallel application is mapped to the proposed
    architecture in Section~\ref{sec:mappings}, emphasizing the usage of the architectural extensions proposed in this
    work. We also show how the convolutional kernels are combined to implement a full neural network and how size
    mismatches can be handled without significant overhead.
    
  \item Section~\ref{sec:experiments} presents an analysis of the gains obtained by our innovations compared to the
    mentioned baseline architectures. This is performed for the metrics relevant to memory bandwidth (which is our main
    focus in this paper) such as memory accesses, latency, PE utilization, among others.
\end{itemize}

%% file: parts/background.tex
\section{Background and state-of-the-art summary}\label{sec:background}
This section describes the state-of-the-art architectures that have become prominent actors in the ML/AI domain:
systolic arrays, vector processors and GPUs. We primarily focus on their memory hierarchies and data movement. The next
section will describe the limitations of these architectures and motivate the introduction of a new memory hierarchy to
overcome some of these limitations.

\newlength{\twosubht}
\newsavebox{\twosubbox}
\begin{figure}[tb]
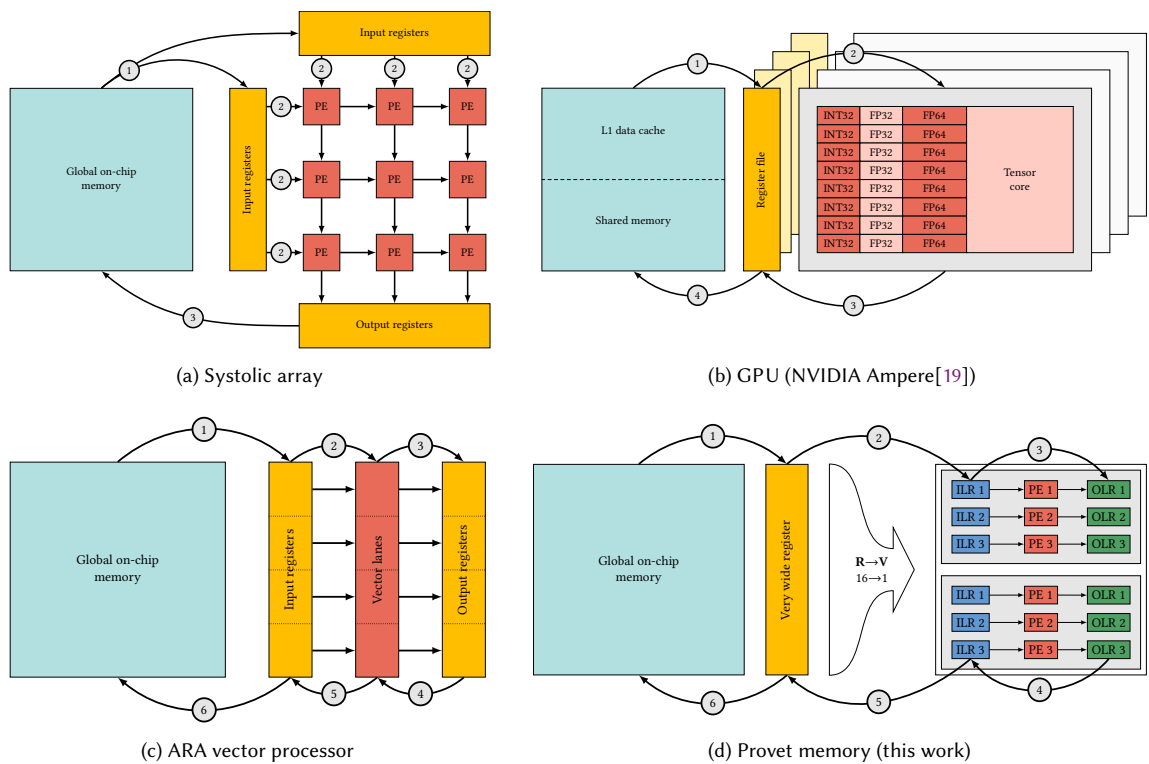

    \sbox\twosubbox{%
      \resizebox{\dimexpr\textwidth-2em}{!}{%
        \includestandalone[height=4.0cm]{./tikz/gpu_hierarchy}%
        \includestandalone[height=4.0cm]{./tikz/sa_hierarchy}%
      }%
    }
    \setlength{\twosubht}{\ht\twosubbox}
    \centering
    \subcaptionbox{Systolic array\label{fig:sa_hierarchy}}{%
        \includestandalone[height=\twosubht]{./tikz/sa_hierarchy}%
    }\hfill%
    \subcaptionbox{GPU (NVIDIA Ampere\cite{ampere2020})\label{fig:gpu_hierarchy}}{%
        \includestandalone[height=\twosubht]{./tikz/gpu_hierarchy}%
    }\vspace{1em}
    %
    \sbox\twosubbox{%
      \resizebox{\dimexpr\textwidth-2em}{!}{%
        \includestandalone[height=4.0cm]{./tikz/ara_hierarchy}%
        \includestandalone[height=4.0cm]{./tikz/provet_hierarchy}%
      }%
    }
    \setlength{\twosubht}{\ht\twosubbox}
    \centering
    \subcaptionbox{ARA vector processor\label{fig:ara_hierarchy}}{%
        \includestandalone[height=\twosubht]{./tikz/ara_hierarchy}%
    }\hfill%
    \subcaptionbox{Provet memory (this work)\label{fig:provet_hierarchy}}{%
        \includestandalone[height=\twosubht]{./tikz/provet_hierarchy}%
    }\vspace{1em}
    \caption{Comparison of the memory hierarchies. The numbered white arrows show the possible paths data can take in
    the architecture when moving up and down the memory hierarchy through the processing elements (PEs)}
    \label{fig:smemory_hierarchies}
    \Description{Comparison of the memory hierarchies. The numbered white arrows show the possible paths data can take in
    the architecture when moving up and down the memory hierarchy through the processing elements (PEs)}
\end{figure}

Figure~\ref{fig:smemory_hierarchies} illustrates the memory hierarchies for (\ref{fig:sa_hierarchy})~systolic arrays,
(\ref{fig:ara_hierarchy})~a general vector processor (ARA~\cite{Perotti2022}), (\ref{fig:gpu_hierarchy})~a GPGPU (NVIDIA
Ampere~\cite{ampere2020}), and (\ref{fig:provet_hierarchy})~our proposed solution, which we refer to as \emph{Provet}
hereafter. We compare these three architecture classes according to: (1) how the Processing Elements (PEs) are
organized, (2) their memory hierarchy, and (3) their interconnect. The descriptions of systolic arrays draw on MIT's
Eyeriss~\cite{Chen2016} and Google's TPU~\cite{Jouppi2017}, those of vector processors on PULP's ARA~\cite{Perotti2022},
and those of GPUs on NVIDIA Ampere~\cite{ampere2020}.

\subsection{Systolic arrays}
Systolic arrays arrange PEs in a 2D grid. Their memory hierarchy is typically straightforward, consisting of an input and
an output register that supply data to, and collect results from, the array. The bandwidth between these registers and
the array is limited by the number of PEs situated on the array’s boundary, which scales proportionally to the square
root of the total number of PEs. The interconnect usually employs nearest-neighbor links, which simplifies hardware costs
but constrains the range of possible data transfers inside the array.

Because of this layout, the organization of PEs restricts the scalability of the memory bandwidth
(Section~\ref{sec:bandwidth-scalability}) and makes the array susceptible to mismatches in size
(Section~\ref{sec:size-mismatch}). Finally, limitations in the interconnect further constrain the kinds of dataflows that
can be efficiently mapped onto the array (Section~\ref{sec:interconnect-limitations}).

\subsection{Vector processors}
Conventional vector processors place PEs in a one-dimensional arrangement, with each PE connected to its own local
memory (input and output registers). Data movement within a single vector lane is confined to the slice of memory
allocated to that lane. Their memory hierarchy is much like that of systolic arrays, featuring an input and an output
register to feed data to the array and store outcomes. However, the interconnect within a vector lane is more adaptable,
allowing shuffling or permutation among PEs. On the other hand, inter-lane communication can only happen through a
shared global interconnect.

Vector processors also experience issues stemming from size mismatches (Section~\ref{sec:size-mismatch}) and are subject
to interconnect limitations (Section~\ref{sec:interconnect-limitations}).

\subsection{GPUs}
GPUs organize their PEs in multiple Streaming Multiprocessors (SMs) (4 PEs per SM in NVIDIA's Ampere architecture 
\cite{ampere2020}). Each PE has its own register file, and PEs within the same SM can access a shared L1 cache and/or a
scratchpad (i.e., shared memory). Multiple SMs share a large L2 cache. Accessing data across SMs require traversing the
higher latency L2 cache and/or main memory (i.e., HBM memory in Ampere).
The memory hierarchy of GPUs is deep, with multiple levels of caches and scratchpads to enable both fast access to small
caches and huge storage capacity for the big workloads of modern applications. The interconnect is more flexible than
systolic arrays and vector processors, with multiple levels of caches and scratchpads that allow for more complex data
movement patterns. However, the interconnect is still limited to the PEs within the same SM, and accessing data across
SMs requires traversing the higher-latency L2 cache and/or main memory.

Similar to vector processors, GPUs are also sensitive to size mismatches, described in Section \ref{sec:size-mismatch},
and the interconnect limitations, described in Section \ref{sec:interconnect-limitations}.

%% file: tikz/gpu_hierarchy.tex
\begin{tikzpicture}

  \node [draw, fill=light_turquoise, rectangle, minimum width=5cm, minimum height=5cm, align=center]
    (A) at (0,0) {};

  \node [draw, anchor=west, fill=light_yellow, rectangle, minimum width=1cm, minimum height=5cm]
    at ($(A.east)+(1.8,1.5)$) (B) {};
  \node [draw, anchor=west, fill=light_yellow, rectangle, minimum width=1cm, minimum height=5cm]
    at ($(A.east)+(1.3,1)$) (B) {};
  \node [draw, anchor=west, fill=light_yellow, rectangle, minimum width=1cm, minimum height=5cm]
    at ($(A.east)+(0.8,0.5)$) (B) {};
  \node [draw, anchor=west, fill=yellow, rectangle, minimum width=1cm, minimum height=5cm]
    at ($(A.east)+(0.5,0)$) (B) {};

  \node [fit=(A.north west) (A.east)] {L1 data cache};
  \node [fit=(A.south west) (A.east)] {Shared memory};
  \draw [densely dashed] (A.east) -- (A.west);

  \node [rotate=90]  at (B) {Register file};

  \pgfmathsetmacro{\boxheight}{(5 - 1)/8}
  \pgfmathsetmacro{\boxwidth}{(8-1)/6}

  \node [draw, rectangle, fill=light_grey!20, minimum width=8cm, minimum height=5cm, anchor=west] (C) at
    ($(B.east)+(2,1.5)$) {};
  \node [draw, rectangle, fill=light_grey!20, minimum width=8cm, minimum height=5cm, anchor=west] (C) at
    ($(B.east)+(1.5,1)$) {};
  \node [draw, rectangle, fill=light_grey!20, minimum width=8cm, minimum height=5cm, anchor=west] (C) at
    ($(B.east)+(1,0.5)$) {};
  \node [draw, rectangle, fill=light_grey, minimum width=8cm, minimum height=5cm, anchor=west] (C) at
    ($(B.east)+(0.5,0)$) {};
  \foreach \i in {0,1,2,3,4,5,6,7}{
    \node [rectangle, draw, fill=brick, minimum width=\boxwidth cm, minimum height=\boxheight cm, anchor=north west] at ($(C.north west) + (0.5,-0.5) + (0,-\i*\boxheight)$) {INT32};
    \node [rectangle, draw, fill=light_brick, minimum width=\boxwidth cm, minimum height=\boxheight cm, anchor=north west] at ($(C.north west) + (0.5,-0.5) + (\boxwidth,-\i*\boxheight)$) {FP32};
    \node [rectangle, draw, fill=brick, minimum width=1.5*\boxwidth cm, minimum height=\boxheight cm, anchor=north
      west] at ($(C.north west) + (0.5,-0.5) + (2*\boxwidth,-\i*\boxheight)$) {FP64};
  }
  \node [rectangle, draw, fill=light_brick, minimum width=2.5*\boxwidth cm, minimum height=8*\boxheight cm, anchor=north west, align=center] at ($(C.north west) + (0.5,-0.5) + (3.5*\boxwidth,0)$) {Tensor\\core};
  
  \draw[very thick, -latex] (A.north) to[out=40, in=140] node[pos=0.5, draw, circle, fill=light_grey] {1} (B.north);
  \draw[very thick, -latex] (B.north) to[out=40, in=140] node[pos=0.5, draw, circle, fill=light_grey] {2} (C.north);
  \draw[very thick, -latex] (C.south) to[out=-140, in=-40] node[pos=0.5, draw, circle, fill=light_grey] {3} (B.south);
  \draw[very thick, -latex] (B.south) to[out=-140, in=-40] node[pos=0.5, draw, circle, fill=light_grey] {4} (A.south);

  \coordinate (nw) at ($(C.south west)+(0,-1)$);  
  \coordinate (se) at ($(C.south east)+(0,-2)$);  

  \node[fit=(nw) (se)] (or) {};
  \coordinate (nw) at ($(C.north west)+(0,1)$);  
  \coordinate (se) at ($(C.north east)+(0,2)$);  

  \node[fit=(nw) (se)] (ir2) {};

\end{tikzpicture}

%% file: tikz/sa_hierarchy.tex
\begin{tikzpicture}

  \node [draw, fill=light_turquoise, rectangle, minimum width=5cm, minimum height=5cm, align=center]
    (A) at (0,0) {Global on-chip\\memory};
  \node [draw, anchor=west, fill=yellow, rectangle, minimum width=1cm, minimum height=5cm]
    at ($(A.east)+(1,0)$) (B) {};

  \node [rotate=90]  at (B) {Input registers};

  \node [rectangle, minimum width=5cm, minimum height=5cm, anchor=west] (C) at ($(B.east)+(1,0)$) {};

  \node [draw, fill=brick, rectangle, anchor=north west, minimum size=1cm] (pe00) at (C.north west) {PE};
  \node [draw, fill=brick, rectangle, anchor=north, minimum size=1cm] (pe01) at (C.north) {PE};
  \node [draw, fill=brick, rectangle, anchor=north east, minimum size=1cm] (pe02) at (C.north east) {PE};
  \node [draw, fill=brick, rectangle, anchor=west, minimum size=1cm] (pe10) at (C.west) {PE};
  \node [draw, fill=brick, rectangle, minimum size=1cm] (pe11) at (C) {PE};
  \node [draw, fill=brick, rectangle, anchor=east, minimum size=1cm] (pe12) at (C.east) {PE};
  \node [draw, fill=brick, rectangle, anchor=south west, minimum size=1cm] (pe20) at (C.south west) {PE};
  \node [draw, fill=brick, rectangle, anchor=south, minimum size=1cm] (pe21) at (C.south) {PE};
  \node [draw, fill=brick, rectangle, anchor=south east, minimum size=1cm] (pe22) at (C.south east) {PE};
  
  \coordinate (nw) at ($(C.south west)+(0,-1)$);  
  \coordinate (se) at ($(C.south east)+(0,-2)$);  

  \node[draw, fill=yellow, fit=(nw) (se)] (or) {};
  \node at (or) {Output registers};

  \coordinate (nw) at ($(C.north west)+(0,1)$);  
  \coordinate (se) at ($(C.north east)+(0,2)$);  

  \node[draw, fill=yellow, fit=(nw) (se)] (ir2) {};
  \node at (ir2) {Input registers};

  \begin{scope}[very thick, latex-]
    \draw (pe00.west) -- node[pos=0.6, draw, circle, fill=light_grey] {2} (pe00.west-|B.east);
    \draw (pe01.west) -- (pe00.east);
    \draw (pe02.west) -- (pe01.east);

    \draw (pe10.west) -- node[pos=0.6, draw, circle, fill=light_grey] {2} (B.east);
    \draw (pe11.west) -- (pe10.east);
    \draw (pe12.west) -- (pe11.east);

    \draw (pe20.west) -- node[pos=0.6, draw, circle, fill=light_grey] {2} (pe20.west-|B.east);
    \draw (pe21.west) -- (pe20.east);
    \draw (pe22.west) -- (pe21.east);

    \draw (pe00.north) -- node[pos=0.6, draw, circle, fill=light_grey] {2} (ir2.south-|pe00.north);
    \draw (pe01.north) -- node[pos=0.6, draw, circle, fill=light_grey] {2} (ir2.south);
    \draw (pe02.north) -- node[pos=0.6, draw, circle, fill=light_grey] {2} (ir2.south-|pe02.north);

    \draw (pe10.north) -- (pe00.south);
    \draw (pe11.north) -- (pe01.south);
    \draw (pe12.north) -- (pe02.south);

    \draw (pe20.north) -- (pe10.south);
    \draw (pe21.north) -- (pe11.south);
    \draw (pe22.north) -- (pe12.south);

    \draw (or.north-|pe20.south) -- (pe20.south);
    \draw (or.north-|pe21.south) -- (pe21.south);
    \draw (or.north-|pe22.south) -- (pe22.south);
  \end{scope}

  \draw[very thick, -latex] (or.west) to[out=180, in=-40] node[pos=0.5, draw, circle, fill=light_grey] {3} (A.south);
  \draw[very thick, -latex] (A.north) to[out=40, in=180] (ir2.west);
  \draw[very thick, -latex] (A.north) to[out=40, in=140] node[pos=0.2, draw, circle, fill=light_grey] {1} (B.north);

\end{tikzpicture}

%% file: tikz/ara_hierarchy.tex
\begin{tikzpicture}

  \node [draw, fill=light_turquoise, rectangle, minimum width=5cm, minimum height=5cm, align=center]
    (A) at (0,0) {Global on-chip\\memory};
  \node [draw, anchor=west, fill=yellow, rectangle, minimum width=1cm, minimum height=5cm]
    at ($(A.east)+(1,0)$) (B) {};

  \node [rotate=90]  at (B) {Input registers};

  \node [rectangle, draw, fill=brick, minimum width=1cm, minimum height=5cm, anchor=west] (C) at ($(B.east)+(1,0)$) {};
  \node [rotate=90] at (C) {Vector lanes};

  \node [rectangle, draw, fill=yellow, minimum width=1cm, minimum height=5cm, anchor=west] (D) at ($(C.east)+(1,0)$) {};
  \node [rotate=90] at (D) {Output registers};

  \pgfmathsetmacro{\dist}{5/4}
  \foreach \i in {1,2,3}{
    \draw[densely dotted] ($(B.south west) + (0,\i*\dist)$) --++ (1,0);
    \draw[densely dotted] ($(C.south west) + (0,\i*\dist)$) --++ (1,0);
    \draw[densely dotted] ($(D.south west) + (0,\i*\dist)$) --++ (1,0);
  }
  \foreach \i in {1,2,3,4}{
    \draw[very thick, -latex] ($(B.south east) + (0,\i*\dist) - (0,0.5*\dist)$) --++ (1,0);
    \draw[very thick, -latex] ($(C.south east) + (0,\i*\dist) - (0,0.5*\dist)$) --++ (1,0);
  }

  \draw[very thick, -latex] (A.north) to[out=40, in=140] node[pos=0.5, draw, circle, fill=light_grey] {1} (B.north);
  \draw[very thick, -latex] (B.north) to[out=40, in=140] node[pos=0.5, draw, circle, fill=light_grey] {2} (C.north);
  \draw[very thick, -latex] (C.north) to[out=40, in=140] node[pos=0.5, draw, circle, fill=light_grey] {3} (D.north);
  \draw[very thick, -latex] (D.south) to[out=-140, in=-40] node[pos=0.5, draw, circle, fill=light_grey] {4} (C.south);
  \draw[very thick, -latex] (C.south) to[out=-140, in=-40] node[pos=0.5, draw, circle, fill=light_grey] {5} (B.south);
  \draw[very thick, -latex] (B.south) to[out=-140, in=-40] node[pos=0.5, draw, circle, fill=light_grey] {6} (A.south);

\end{tikzpicture}

%% file: tikz/provet_hierarchy.tex
\begin{tikzpicture}

  \node [draw, fill=light_turquoise, rectangle, minimum width=5cm, minimum height=5cm, align=center]
    (A) at (0,0) {Global on-chip\\memory};
  \node [draw, anchor=west, fill=yellow, rectangle, minimum width=1cm, minimum height=5cm]
    at ($(A.east)+(0.5,0)$) (B) {};

  \node [rotate=90]  at (B) {Very wide register};

  \node [rectangle, minimum width=2cm, minimum height=5cm, anchor=west] (C) at ($(B.east)+(0.5,0)$) {};

  \draw (C.north west) to [out=0, in=180] ($(C.east)+(-0.5,0.5)$) --++(0,0.5) -- (C.east)
    --($(C.east)+(-0.5,-1)$) --++(0,0.5) to [out=180, in=0] (C.south west) -- cycle;

  \node[align=center] at (C) {\textbf R$\rightarrow$\textbf V \\ 16$\rightarrow$1};

  \node [draw, rectangle, minimum width=5cm, minimum height=5cm, anchor=west] (D) at ($(C.east)+(0.5,0)$) {};

  \coordinate (nw) at ($(D.north west)+(0.25,-0.25)$);  
  \coordinate (se) at ($(D.east)+(-0.25,0.25)$);        
  \node[draw, fill=light_grey, fit=(nw) (se)] (pe1) {};
  \coordinate (nw) at ($(D.south west)+(0.25,0.25)$);  
  \coordinate (se) at ($(D.east)+(-0.25,-0.25)$);        
  \node[draw, fill=light_grey, fit=(nw) (se)] (pe2) {};


  \node [draw, fill=sky_blue, rectangle, anchor=north west] at ($(pe1.north west)+(0.25,-0.25)$) (ilr0) {ILR 1};
  \node [draw, fill=sky_blue, rectangle, anchor=west] at ($(pe1.west)+(0.25,0)$) (ilr1) {ILR 2};
  \node [draw, fill=sky_blue, rectangle, anchor=south west] at ($(pe1.south west)+(0.25,0.25)$) (ilr2) {ILR 3};

  \node [draw, fill=brick, rectangle, anchor=north] at ($(pe1.north)+(0,-0.25)$) (ppe1) {PE 1};
  \node [draw, fill=brick, rectangle] at (pe1) (ppe2) {PE 2};
  \node [draw, fill=brick, rectangle, anchor=south] at ($(pe1.south)+(0,0.25)$) (ppe3) {PE 3};

  \node [draw, fill=green, rectangle, anchor=north east] at ($(pe1.north east)+(-0.25,-0.25)$) (olr0) {OLR 1};
  \node [draw, fill=green, rectangle, anchor=east] at ($(pe1.east)+(-0.25,0)$) (olr1) {OLR 2};
  \node [draw, fill=green, rectangle, anchor=south east] at ($(pe1.south east)+(-0.25,0.25)$) (olr2) {OLR 3};

  \draw [-latex] (ilr0.east) -- (ppe1.west);
  \draw [-latex] (ilr1.east) -- (ppe2.west);
  \draw [-latex] (ilr2.east) -- (ppe3.west);
  \draw [-latex] (ppe1.east) -- (olr0.west);
  \draw [-latex] (ppe2.east) -- (olr1.west);
  \draw [-latex] (ppe3.east) -- (olr2.west);

  \node [draw, fill=sky_blue, rectangle, anchor=north west] at ($(pe2.north west)+(0.25,-0.25)$) (ilr1) {ILR 1};
  \node [draw, fill=sky_blue, rectangle, anchor=west] at ($(pe2.west)+(0.25,0)$) (ilr2) {ILR 2};
  \node [draw, fill=sky_blue, rectangle, anchor=south west] at ($(pe2.south west)+(0.25,0.25)$) (ilr3) {ILR 3};

  \node [draw, fill=brick, rectangle, anchor=north] at ($(pe2.north)+(0,-0.25)$) (ppe1) {PE 1};
  \node [draw, fill=brick, rectangle] at (pe2) (ppe2) {PE 2};
  \node [draw, fill=brick, rectangle, anchor=south] at ($(pe2.south)+(0,0.25)$) (ppe3) {PE 3};

  \node [draw, fill=green, rectangle, anchor=north east] at ($(pe2.north east)+(-0.25,-0.25)$) (olr1) {OLR 1};
  \node [draw, fill=green, rectangle, anchor=east] at ($(pe2.east)+(-0.25,0)$) (olr2) {OLR 2};
  \node [draw, fill=green, rectangle, anchor=south east] at ($(pe2.south east)+(-0.25,0.25)$) (olr3) {OLR 3};

  \draw [-latex] (ilr1.east) -- (ppe1.west);
  \draw [-latex] (ilr2.east) -- (ppe2.west);
  \draw [-latex] (ilr3.east) -- (ppe3.west);
  \draw [-latex] (ppe1.east) -- (olr1.west);
  \draw [-latex] (ppe2.east) -- (olr2.west);
  \draw [-latex] (ppe3.east) -- (olr3.west);

  \draw [very thick, -latex] (A.north) to[out=40, in=140] node[pos=0.5, draw, circle, fill=light_grey] {1} (B.north);
  \draw [very thick, -latex] (B.north) to[out=40, in=140] node[pos=0.5, draw, circle, fill=light_grey] {2} (ilr0.north);
  \draw [very thick, -latex] (ilr0.north) to[out=50, in=130] node[pos=0.5, draw, circle, fill=light_grey] {3} (olr0.north);
  \draw [very thick, -latex] (olr3.south) to[out=-130, in=-50] node[pos=0.5, draw, circle, fill=light_grey] {4} (ilr3.south);
  \draw [very thick, -latex] (ilr3.south) to[out=-140, in=-40] node[pos=0.5, draw, circle, fill=light_grey] {5} (B.south);
  \draw [very thick, -latex] (B.south) to[out=220, in=320] node[pos=0.5, draw, circle, fill=light_grey] {6} (A.south);

\end{tikzpicture}

%% file: parts/motivations.tex
\section{Motivations}\label{sec:motivations}
Based on the state-of-the-art architectures described in the previous section, we identify four key limitations that
motivate the introduction of a new memory hierarchy. The first two primarily affect systolic arrays, while the latter
are common to both systolic arrays, GPUs and vector processors.

\subsection{Memory bandwidth scalability}\label{sec:bandwidth-scalability}
Array (2D) organizations intrinsically limit the scalability of the architecture in terms of memory bandwidth. The input
and output bandwidth of the array scales with the square root of the number of processing elements, which means that the
size of the array required to achieve a given bandwidth grows quadratically. This, coupled with the size mismatch
(sec.~\ref{sec:size-mismatch}) and the interconnect limitations (sec.~\ref{sec:interconnect-limitations}) limits the
size of the array that can be used in practice. Thus, the memory bottleneck can only be mitigated by exploiting the data
reuse within the array, which is not always possible in all applications. We show in section~\ref{sec:experiments} how
some networks which exhibit little data reuse, like MobileNet or SqueezeNet, lead to a significant underutilization in
array-based architectures (e.g., Eyeriss, TPU).

This limitation is addressed in section~\ref{sec:memory_bandwidth_comparison}.

\subsection{Size mismatch sensitivity}\label{sec:size-mismatch}
Increasing the size of the arrays to achieve higher memory bandwidth can result in arrays that are too large to
naturally fit the dimensions of the application kernels. Complex dataflows and folding methods are required to map the
application to the array to fully utilize the array. Even when using automatic dataflow design-space exploration
tools~\cite{9360462, 8695666}, the rigid interconnect of the array can limit the options and force the data to jump over
multiple PEs, which increases the overhead and latency. This issue is further exacerbated with the scaling of the array
size. Linear (1D) organization of the processing elements can mitigate this problem as data can be linearly spatially
distributed in the memory, which allows for a direct path between memory and PEs, removing the jumps and not limiting the
scalability of the array. The key difference between linear (1D) vs array (2D) organizations is that data can only be fed
to the edge PEs of the arrays, while linear organizations can deliver data to all the PEs. Refer to
Figure~\ref{fig:smemory_hierarchies}.

This limitation is addressed in section~\ref{sec:interconnect_comparison}.

\subsection{Interconnect limitations}\label{sec:interconnect-limitations}
Interconnects in both systolic arrays, vector processors and GPUs are simple and rigid. This is a conscious design
choice to minimize the overhead and complexity; however, it can limit the flexibility and restrict the operations that
can be naturally mapped to the array. For the relevant case of convolutional kernels, the sliding nature of the kernel
cannot be naturally mapped without having to use im2col transformations or its derivatives.
Such transformations are required to exploit the efficient GEneral Matrix Multiplication (GEMM) kernels present in
commonly used linear algebra libraries, such as NVIDIA cuBLAS \cite{NVIDIA_cuBLAS}. However, im2col introduces large
data redundancy. For example, a $7\times7$ kernel with a stride of 1 convolving a $256\times256$ image requires a
$7^2\times256^2$ matrix, going from 65 thousand elements to 3 million elements, a $\times46$ increase. Furthermore,
the large size difference between the matrices can lead to performance decreases
\cite{zhang2018highperformancezeromemoryoverhead, 10.1007/3-540-45545-0_15} and the overhead of the
transformation can be significant for small kernels. Even when using more efficient variations of the im2col
transformation, such as implicit GEMM, can lower the memory overhead by one order of magnitude, the overhead will always
be higher than a factor of two (100\%) \cite{Zhou2021}.

This limitation is addressed in section~\ref{sec:interconnect}, where we show how a more flexible interconnect can
allow for more complex data movement patterns that support the sliding nature of the convolutional kernel.

\subsection{Dependency on data reuse}\label{sec:data-reuse}
Systolic arrays, vector processors and GPUs rely heavily on data reuse to reduce the memory accesses and maintain a high
utilization of the processing elements. However, not all ML and AI workloads exhibit high data reuse, which could lead
to underutilization of the array and suboptimal performance. A common example in CNN are depth-wise separable
convolutions, found in the MobileNet class CNNs \cite{DBLP:journals/corr/HowardZCKWWAA17}.

We show in section \ref{sec:experiments} how our proposal does not rely on data reuse to mitigate the memory bottleneck
challenges, although data reuse greatly benefits that aspect, even in applications with little data reuse, the design is
not memory-throttled. Any additional data reuse will enhance the performance and reduce the memory accesses on top of
the obtained by the architectural gains.

%% file: parts/architecture.tex
\section{Detailed architectural description}\label{sec:detailed_description} 
In this section, we present a vector architectural template that mitigates the limitations of current architectures
based on systolic (2D) and vector (1D) arrays. The proposed architecture, hereafter referred to as Provet, is based on a
1D organization, however it introduces extra elements in the memory hierarchy and interconnect making it significantly
different from traditional vector processors or GPUs. This section describes Provet's architecture in detail, including
the memory hierarchy, components, and control structure. The next section (sec.~\ref{sec:comparison}) compares the
proposed architecture with others, highlighting key differences. While it borrows many concepts, such as the linear (1D)
organization, the added elements introduce significant improvements and distinctions which we detail in this section. 

\subsection{Memory hierarchy}\label{sec:memory_hierarchy}
We propose an ultra-wide and shallow memory as the main global on-chip memory. An ultra-wide memory inherently has a
very high bandwidth, while maintaining a low energy consumption. This can be understood by observing that the energy
cost of accessing an SRAM consists of two main components: width-dependent and depth-dependent. The analysis is
targeted to Provet SRAM memory because this paper focuses only on on-chip aspects of the architecture. Although the same
conclusions can be drawn for off-chip memories, the analysis for the implications on the interconnect of off-chip
memories falls out of the scope of this paper.

\begin{figure}[tb]
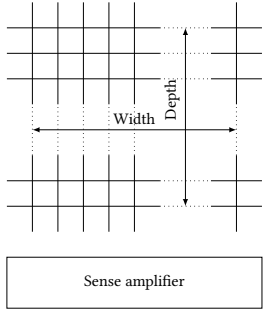
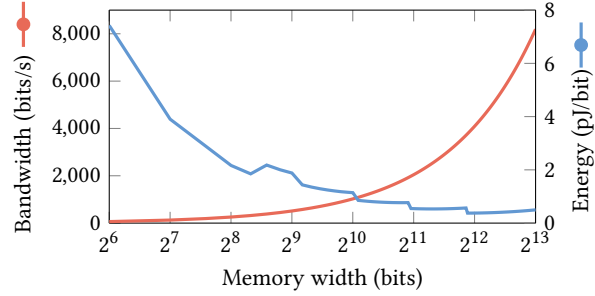

  \centering
  \hspace*{\fill}%
  \begin{subfigure}[c]{0.25\textwidth}
    \centering
    \includestandalone[width=0.9\linewidth]{./tikz/sram_structure}
    \caption{Simplified structure of an SRAM memory.}
    \label{fig:sram_structure}
    \Description{Simplified structure of an SRAM memory.}
  \end{subfigure}%
  \hspace*{\fill}%
  \begin{subfigure}[c]{0.75\textwidth}
    \centering
    \includestandalone{./tikz/sram_energy}
    \caption{Energy cost of accessing an SRAM with different widths and depths.}
    \label{fig:sram_energy}
    \Description{Energy cost of accessing an SRAM with different widths and depths.}
  \end{subfigure}
  \caption{Simplified SRAM memory structure and energy cost estimations.}
  \hspace*{\fill}%
\end{figure}

As illustrated in Figure~\ref{fig:sram_structure}, the length of the word lines is proportional to the number of bit
lines\textit{ W}, while the length of the bit lines is proportional to the depth of the memory \textit{D}, representing
the number of word lines. Let the energy cost per unit length of the bit and word lines be denoted as\textit{ BL} and
\textit{WL}, respectively. Consequently, the cost of energizing a single bit line and word line would $D\cdot
$\textit{BL} and $W \cdot $\textit{WL} respectively.  To access one word requires energizing all \textit{W} bit lines,
and one word line as expressed below: \begin{equation} \overbrace{W \cdot D \cdot BL}^{\text{energize}\ W\
  \text{bitlines}} + \underbrace{1 \cdot W \cdot WL}_{{\text{energize}\ 1\ \text{wordline}}} \end{equation}

To gauge the impact of the width of memory on accessing a single word, we normalize the above expression with the width
as shown below. 

\begin{equation} \frac{W \cdot D \cdot BL + 1 \cdot W \cdot WL}{W} = D\cdot BL + WL \end{equation}

This shows that maintaining an ultra-wide and shallow memory provides a high bandwidth while maintaining a low access
cost, compared to an equivalent memory with a less aggressive aspect ratio.

To validate such claims, we have performed a simple simulation of the energy cost of accessing an SRAM with different
widths and depths using the CACTI tool~\cite{Jouppi2012} and present the results in Figure~\ref{fig:sram_energy}. It is
clear that the energy cost per bit decreases with the width of the SRAM, while the bandwidth increases. This validates
the lack of an energy consumption penalty when using an ultra-wide memory.

\begin{figure}[tb]
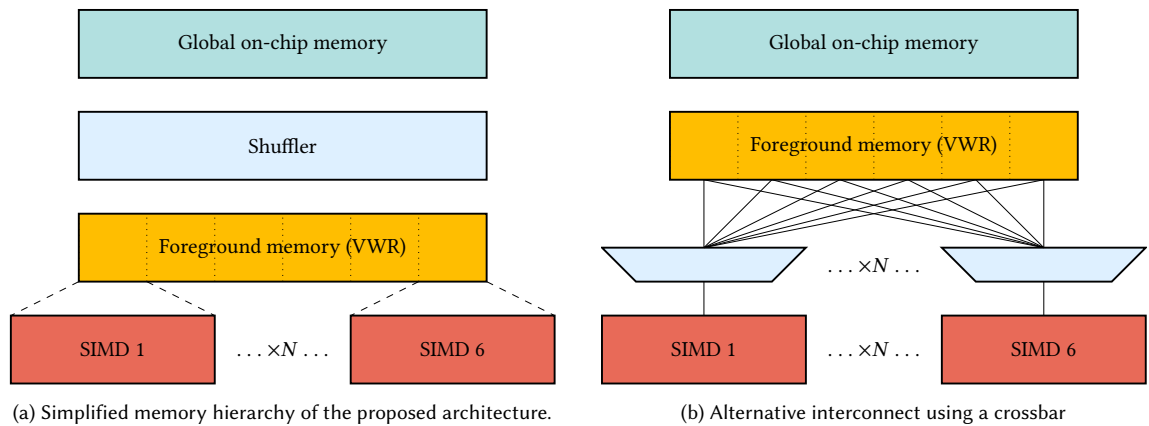

  \centering
  \begin{subfigure}[b]{0.48\textwidth}
    \centering
    \includestandalone[width=\linewidth]{./tikz/memory_hierarchy}
    \caption{Simplified memory hierarchy of the proposed architecture.}
    \label{fig:memory_hierarchy}
    \Description{Simplified memory hierarchy of the proposed architecture.}
  \end{subfigure}%
  \hfill%
  \begin{subfigure}[b]{0.48\textwidth}
    \centering
    \includestandalone[width=\linewidth]{./tikz/memory_hierarchy_alt}
    \caption{Alternative interconnect using a crossbar}
    \label{fig:memory_hierarchy_alt}
    \Description{Alternative interconnect using a crossbar}
  \end{subfigure}
  \caption{Memory hierarchy and interconnect of two implementation variants of the proposed architecture.}
  \label{fig:memory_hierarchy_and_architecture}
\end{figure}


While the ultra-wide memory provides a low energy access cost, accessing it repeatedly for each computation would still
cause a significant energy overhead. To mitigate this, we introduce a foreground memory (see
Fig.~\ref{fig:memory_hierarchy}) in the form of a very wide register (VWR) that acts as a buffer between the ultra-wide
memory and the vector SIMD units. This register is a single (ultra-wide) word register and \emph{not} a multi-port
register file. This means that it does not have an address space thus it does not have the overhead of the address
decoding and the multiplexing of the data (which happens in a multi-port register file). Furthermore, the VWR has an
asymmetric interface that matches the width of the ultra-wide memory and the SIMD units, on the other hand, as shown in
Figure~\ref{fig:memory_hierarchy}. 
The assymetry in the VWR provides an alternative mechanism to reduce the amount of memory accesses on top of the inherent 
data reuse in the application. In a general case, the number of accesses to the VWR will be $N$ times higher than the
number of accesses to the SRAM memory, where $N$ is the width ratio between the SRAM width and the SIMD width.
Any data reuse present in the application will further reduce the number of accesses to the SRAM memory. The experiment
results presented in section~\ref{sec:experiments} validate this claim by showing a significant improvement in the
compute-to-memory access ration in situations where other architectures heavily throttle due to limited data reuse.

%

\subsection{Interconnect}\label{sec:interconnect}
The interconnect between the VWR and SIMD units is pitch-aligned to simplify the layout and create a regular structure.
This means that each SIMD unit can access only a specific slice of the VWR. To allow SIMD units to access data that is
not aligned with their slice, we introduce a shuffler. This shuffler is positioned between the SRAM and the VWR and
it can be customized to provide different shuffling patterns, granularities, and ranges. The shuffling operations
required are determined by profiling the application or application domain. 

\begin{table}[tb]
  \centering
  \caption{Comparison of the area of the proposed shuffler and a generic crossbar implementation.}
  \label{tab:shuffler_area}
  \begin{tabular}{lrrr}
    \toprule
    & \textbf{Shuffler} & \textbf{Crossbar} & \textbf{Difference} \\
    \midrule
    \textbf{Area} (mm\textsuperscript{2}) & 0.13 & 0.88 & $\times$6.82 \\
    \textbf{Gate count} & 16k & 86k & $\times$5.38 \\
    \textbf{Wire length} (mm) & 4.3 & 33.1 & $\times$7.67 \\
    \bottomrule
  \end{tabular}
\end{table}

This interconnect choice is a conscious design decision to minimize the overhead and complexity that could result in a
severe area and energy penalty. Even though the energy analysis is not the focus of this work, we have performed a
simple comparison of the post-layout area of the proposed shuffler and a generic crossbar implementation (see 
Fig.~\ref{fig:memory_hierarchy_alt}) and present the results in Table~\ref{tab:shuffler_area}. The results show that the
shuffler has a significantly smaller footprint in terms of area, gate count and wire length. However, we show in 
section~\ref{sec:example} that the simple shuffler can still provide the necessary flexibility to implement complete CNN
layers without any performance penalty.

The connections inside the SIMD units are a bit more flexible. We introduce multiplexers that allow the data to be moved
between the VFU (vector arithmetic unit) and the local registers. This allows the local registers to be used as
temporary storage for operations like accumulation. The specific details of the individual components of the
architecture are presented in the next section.

\subsection{Detailed architecture description}
\begin{figure}[t]
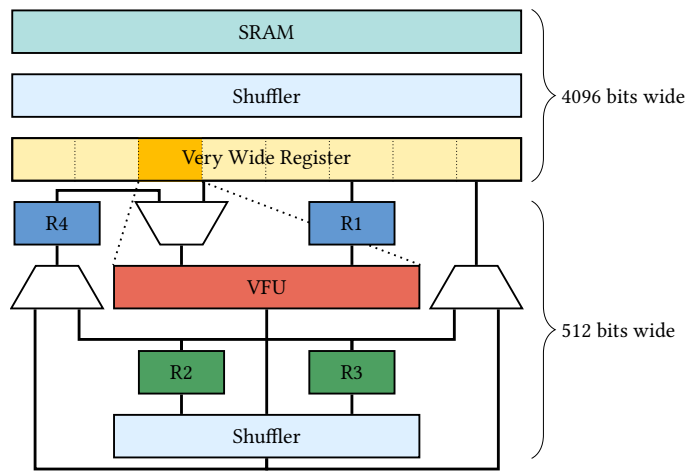

    \centering
    \includestandalone[width=0.6\linewidth]{./tikz/provet_architecture}
    \caption{Proposed architecture overview. The ultra-wide elements (4096 bits) and the SIMD unit (512 bits) are not
      drawn to scale to simplify the diagram.}
    \label{fig:architecture_overview}
    \Description{Provet architecture}
\end{figure}

A detailed diagram of the architecture can be seen in Figure~\ref{fig:architecture_overview}. We now present the elements seen in the figure.

\subsubsection{Ultra-wide shallow SRAM memory} the bulk of the data that will be fetched in the short-term future by the
computation elements. To leverage the properties explained in~\ref{sec:memory_hierarchy} the width of the SRAM is \(8\times\)
bigger than the size of the SIMD unit, however, the depth of the SRAM is small, in the order of 1--32 words.

\subsubsection{Wide coarse-grained tile shuffler}\label{sec:wide_shuffler} is located in between the SRAM and the VWRs.
The shuffler implements the data rearrangement concept explained in~\ref{sec:data_rearrangement}. The tile shuffler is
responsible for the coarse shuffling operations, i.e. it moves large blocks of data but the granularity of the shuffling
distances is coarse. For example, the tile shuffler operates in blocks of data with the same size as the VWR blocks (for
eg. 512 bits) and it moves the block in steps of this same size (also 512 bits). The shuffler can be customized with
different shuffling capabilities, depending on the application needs.

\subsubsection{Narrow file-grained VFU shuffler} is located between the VWR and the SIMD units. The shuffler is
responsible for the fine shuffling operations, i.e. it moves small blocks of data within a SIMD word.
There are two dimensions of the shuffler. One is the size of the blocks it moves or shuffles. As stated above, this is
decided by the aggregate width of VFU ports. The other dimension is the maximum range in terms of steps; each step is
equal to the size of the blocks. A large max range will make the shuffler expensive and inefficient. This dimension
is decided by profiling the intended application or application suite to find out the most widely used range of steps.
Note that it is possible to go beyond the max range by using multiple steps, at the cost of more cycles.

\subsubsection{Very Wide Register (VWR)} is an single-row memory buffer placed between the SRAM and the VFUs. Usually 2
VWRs are allocated to enable concurrent reads and writes to each of these. But even solutions with a single VWR can
function well on condition that the mapping is carefully optimized to accommodate this restriction. The VWR takes to the
extreme the concept explained in~\ref{sec:ultra_wide_mem}. The depth of the VWR is 1, and the width is matched to the
SRAM width. The VWR also has asymmetric port sizes that match the widths of the SRAM and the VFUs. The ratio between the
VWR ports translated into at least a similar ratio of memory accesses. In practice, thanks to the typically big data
reuse opportunities in machine learning computations the memory access ratios are much bigger. This is exemplified in
section~\ref{sec:example} with an illustrative example of a CONV layer. Still, as already mentioned some ML topologies
don't exhibit this large data reuse and they are the most suited candidates for mapping to Provet.

The VWRs are meant to be pitch aligned with the VFUs for a simpler layout, which implies that only the VFU that
physically aligns with a portion of data in the VWR can access it. The tile shuffler (sec.~\ref{sec:tile_shuffler}) is
used to allow VFUs to access data that is not specifically aligned to them. Given the simplified interconnection between
the VWRs and DPUs, the cost of accessing the VWR data is significantly lower than SRAM. The VWRs are used to store the
data used in the inner loops of the application. For each transaction between the VWR and the SRAM, the VWR will serve
data to the DPUs for several cycles. Ideally, this ratio should match the ratio between the VWR port sizes. However, due
to limitations for the data arrangement, it will typically be lower. If the subwords in that data need to be rearranged
during the loop iteration, the DPU shifter can be used (\ref{sec:DPU_shifter}).

\subsubsection{Local registers} In the proposed architecture, there are four local registers: R1 to R4, as shown in
Figure~\ref{fig:architecture_overview}. These registers are typically used to store the weights of the filter in a
weight-stationary dataflow or the partial sum in an output-stationary dataflow. The size of the registers matches the
width of the VFU.

\subsubsection{Vector Functional Unit (VFU)} is an SIMD (Single Instruction Multiple Data) computation unit. The number
of parallel lanes is parametric but natural values are in the 16--64 range. The width of each operand is typically 8 bits,
resulting in a total width of 128--512 bits. The VFU implements the basic operations needed for ML workflows, such as
addition, multiplication, non-linear functions (tanh, sigmoid). A complete list of the VFU modes can be seen in
Table~\ref{tab:isa}.

All VFU computations take 2 operands as inputs and produce an output (with the exception of the non-linear functions).
One of the inputs of the VFU will come from R1 and the other from either R4 or VWR. The output of the VFU can be: stored
in R2 or R3, sent to R4 or VWR (or both), or sent through the VFU shuffler.

The number of VFUs is parametric and can range from 1 to the width of the SRAM. Increasing the number of VFUs enhances
the parallel capabilities of the architecture. Different VFUs can execute different instructions
simultaneously.

\subsubsection{VFU shuffler} \label{sec:DPU_shifter} complements the tile shuffler by providing fine shuffling
granularity but small ranges. The shuffling range of the VFU shuffler is at most the width of the VFU and the
granularity is the size of one operand. This shuffler is key to implementing operations that include a ``sliding'' of
data, such as CONV, MAXPOOL, and AVGPOOL layers, among others.

The shuffler operates from the output of the VFU or R2/R3 registers. Additionally, it can operate directly from the
output of the VWR, bypassing the VFU. This can be useful when data needs to be rearranged inside the VWR but no
computation is needed.

\subsection{Control structure}\label{sec:vfu_instructions}
Even though the control structure of the architecture is not the main focus of this work, for completeness we briefly
decribe the instructions Table~\ref{tab:isa} and DPU modes (VFUX instruction).

\begin{table}[h!]
\caption{Instruction set description}
\label{tab:isa}
\begin{tabular}{p{0.1\linewidth}p{0.8\linewidth}}
    \toprule
    \textbf{Instr.} & \textbf{Description} \\ \midrule
    \textbf{NOP}             & No-operation \\ \midrule
    \multicolumn{2}{c}{Data Transfer Instructions} \\ \midrule
    \textbf{RLB} & Transfers data between SRAM and VWR \\
    \textbf{WLB} & Transfers data between VWR and SRAM \\
    \textbf{VMV} & Transfers data between VWR and local DPU registers \\ \midrule
    \multicolumn{2}{c}{Data Rearrangement Instruction} \\ \midrule
    \textbf{GLMV} & Shuffles the VWR content and store back to itself \\
    \textbf{RMV}  & Shuffles the local DPU register content and stores it to VWR\\
    \textbf{PERM} & Permutes at word-level using the DPU shuffler using a list of (source, destination) pairs that
    represent the movement of the words \\ \midrule
    \multicolumn{2}{c}{Computation Instruction} \\ \midrule
    \textbf{VFUX} & DPU instructions with the following modes: \\
    & Multiply, add, max, multiply-accumulate, add-accumualte, max-accumulate, clip, shift, RELU, sigmoid, tanh  \\
    \textbf{CALC} & Scalar operations using local DPU registers\\ \midrule
    \multicolumn{2}{c}{Control Instruction} \\ \midrule
    \textbf{BRAN} & Branch instruction \\ \bottomrule
\end{tabular}
\end{table}

Due to the very wide nature of this architecture, it can become challenging to effectively control all the resources
without requiring very long control wires that would have a negative impact on the energy efficiency.
To overcome that we propose to decompose each instruction into its ``control actions''. For example, an instruction that
adds R1+VWR and stores it into R2 needs to take the following control actions: configure the VFU mode; enable the VWR,
R1 and R2; configure the multiplexers to select the appropriate inputs. Each of these control actions can be performed
by a distributed control element without the need for a central control unit. We call these distributed elements
\emph{loop buffers}.

The Loop Buffers (LBs) replace the centralized controller that would typically drive the control signals for all components
and instead each loop buffer controls only the component it is associated with. This greatly reduces the length of the
most active control signal wires which have to be activated every cycle when that part of the VFU is operational. Due to
the nature of LB, LB content updates occur much less frequently (10--100\(\times\) less). 

The detailed implementation of the loop buffers and overall control structure falls out of the scope of this paper. We
plan to address all these aspects in future work.

%% file: tikz/sram_structure.tex
\begin{tikzpicture}

  \draw (0,0.5) -- (3,0.5); \draw [dotted] (3,0.5) -- (4,0.5); \draw (4,0.5) -- (5,0.5);
  \draw (0,1) --   (3,1); \draw [dotted] (3,1) -- (4,1); \draw (4,1) -- (5,1);
  \draw (0,3) -- (3,3); \draw [dotted] (3,3) -- (4,3); \draw (4,3) -- (5,3);
  \draw (0,3.5) -- (3,3.5); \draw [dotted] (3,3.5) -- (4,3.5); \draw (4,3.5) -- (5,3.5);
  \draw (0,4) -- (3,4); \draw [dotted] (3,4) -- (4,4); \draw (4,4) -- (5,4);

  \draw (0.5, 0) -- (0.5, 1.5); \draw [dotted] (0.5, 1.5) -- (0.5, 2.5); \draw (0.5, 2.5) -- (0.5, 4.5);
  \draw (1, 0) -- (1, 1.5); \draw [dotted] (1, 1.5) -- (1, 2.5); \draw (1, 2.5) -- (1, 4.5);
  \draw (1.5, 0) -- (1.5, 1.5); \draw [dotted] (1.5, 1.5) -- (1.5, 2.5); \draw (1.5, 2.5) -- (1.5, 4.5);
  \draw (2, 0) -- (2, 1.5); \draw [dotted] (2, 1.5) -- (2, 2.5); \draw (2, 2.5) -- (2, 4.5);
  \draw (2.5, 0) -- (2.5, 1.5); \draw [dotted] (2.5, 1.5) -- (2.5, 2.5); \draw (2.5, 2.5) -- (2.5, 4.5);
  \draw (4.5, 0) -- (4.5, 1.5); \draw [dotted] (4.5, 1.5) -- (4.5, 2.5); \draw (4.5, 2.5) -- (4.5, 4.5);

  \draw (0, -0.5) rectangle (5, -1.5) node [pos=0.5] {Sense amplifier};

  \draw [latex-latex] (0.5, 2) -- (4.5, 2) node [pos=0.5, above] {Width};
  \draw [latex-latex] (3.5,0.5) -- (3.5, 4) node [pos=0.6, above, rotate=90] {Depth};

\end{tikzpicture}

%% file: tikz/sram_energy.tex
  \begin{tikzpicture}
    \pgfplotsset{
        scale only axis,
        scaled x ticks=base 10:3,
        xmin=64, xmax=8192,
        width=0.5\linewidth,
        height=0.25\linewidth,
    }
    \begin{axis}[
      ymin=0, ymax=9000,
      axis y line*=left,
      xlabel={Memory width (bits)},
      ylabel={Bandwidth (bits/s) \ref{plot:bandwidth}},
      xmode=log,
      log basis x=2,
      ]
      \addplot [very thick, color=brick, mark=* ] table [x=bus_width, y=bus_width, col sep=comma, mark=none] {data/energy_vs_width_fit.csv};
      \label{plot:bandwidth}
    \end{axis}
    \begin{axis}[
      ymin=0, ymax=8,
      axis y line*=right,
      hide x axis,
      ylabel={Energy (pJ/bit)\ref{plot:energy}},
      xmode=log,
      log basis x=2,
      legend pos=north west,
      ]
      \addplot [very thick, color=sky_blue, mark=* ] table [x=bus_width, y=energy_per_bit, col sep=comma, mark=none] {data/energy_vs_width_fit.csv};
      \label{plot:energy}
    \end{axis}
  \end{tikzpicture}

%% file: tikz/memory_hierarchy.tex
    \begin{tikzpicture}

      \draw[thick, fill=light_turquoise] (0,4.5) rectangle (6,5.5);
      \node at (3,5) {Global on-chip memory};

      \draw[thick, fill=light_blue] (0,3) rectangle (6,4);
      \node at (3,3.5) {Shuffler};

      \draw[thick, fill=yellow] (0,1.5) rectangle (6,2.5);
      \foreach \x in {0, 1, 2, 3, 4, 5} {
        \draw [dotted] (\x,1.5) -- (\x,2.5); 
      }
      \node at (3,2) {Foreground memory (VWR)};

      \draw[thick, fill=brick] (-1,0) rectangle (2,1);
      \node at (0.5,0.5) {SIMD 1};

      \draw[thick, fill=brick] (4,0) rectangle (7,1);
      \node at (5.5,0.5) {SIMD 6};
      \node at (3, 0.5) {. . . $\times N$ . . .};

      \draw[dashed] (0,1.5) -- (-1,1);  
      \draw[dashed] (1,1.5) -- (2,1);  
      \draw[dashed] (6,1.5) -- (7,1);  
      \draw[dashed] (5,1.5) -- (4,1);  

    \end{tikzpicture}

%% file: tikz/memory_hierarchy_alt.tex
    \begin{tikzpicture}

      \draw[thick, fill=light_turquoise] (0,4.5) rectangle (6,5.5);
      \node at (3,5) {Global on-chip memory};

      \draw[thick, fill=yellow] (0,3) rectangle (6,4);
      \foreach \x in {0, 1, 2, 3, 4, 5} {
        \draw [dotted] (\x,3) -- (\x,4); 
      }
      \node at (3,3.5) {Foreground memory (VWR)};

      \draw[thick, fill=brick] (-1,0) rectangle (2,1);
      \node at (0.5,0.5) {SIMD 1};

      \draw[thick, fill=brick] (4,0) rectangle (7,1);
      \node at (5.5,0.5) {SIMD 6};
      \node at (3, 0.5) {. . . $\times N$ . . .};

      \draw[thick, fill=light_blue] (-0.5,1.5) --++ (-0.5,0.5) --++ (3,0) --++ (-0.5,-0.5) -- cycle;
      \draw (0.5, 2) -- (0.5, 3);
      \draw (0.5, 2) -- (1.5, 3);
      \draw (0.5, 2) -- (2.5, 3);
      \draw (0.5, 2) -- (3.5, 3);
      \draw (0.5, 2) -- (4.5, 3);
      \draw (0.5, 2) -- (5.5, 3);
      \draw (0.5, 1.5) --++ (0,-0.5);

      \node at (3, 1.75) {. . . $\times N$ . . .};
      \draw [thick, fill=light_blue] (6.5, 1.5) --++ (0.5,0.5) --++ (-3,0) --++ (0.5,-0.5) -- cycle;
      \draw (5.5, 2) -- (0.5, 3);
      \draw (5.5, 2) -- (1.5, 3);
      \draw (5.5, 2) -- (2.5, 3);
      \draw (5.5, 2) -- (3.5, 3);
      \draw (5.5, 2) -- (4.5, 3);
      \draw (5.5, 2) -- (5.5, 3);
      \draw (5.5, 1.5) --++ (0,-0.5);

    \end{tikzpicture}

%% file: tikz/provet_architecture.tex
\begin{tikzpicture}
  \path (0,0) node[draw, thick, fill=light_turquoise, rectangle, minimum height=2em, minimum width=0.5\linewidth] (sram) {SRAM};
  \path (sram) ++(0,-3em) node [draw, thick, fill=light_blue, rectangle, minimum height=2em, minimum width=0.5\linewidth]
  (shuffler) {Shuffler};
  \path (shuffler) ++(0,-3em) node [draw, thick, fill=light_yellow, rectangle, minimum height=2em, minimum width=0.5\linewidth]
  (vwr) {};
  \fill [rectangle, fill=yellow] ($(vwr.south west)+(2*0.5*\linewidth/8,0)$) rectangle ($(vwr.north west)+(3*0.5*\linewidth/8,0)$);
  \path (shuffler) ++(0,-3em) node [draw, thick, rectangle, minimum height=2em, minimum width=0.5\linewidth] {Very Wide Register};
  \foreach \x in {1, 2, 3, 4, 5, 6, 7} {
    \draw [densely dotted] ($ (vwr.south west) + (\x*0.5*\linewidth/8,0) $) -- ($ (vwr.north west) + (\x*0.5*\linewidth/8,0) $);
  }

  \path (vwr) ++(0,-6em) node [draw, thick, fill=brick, rectangle, minimum height=2em, minimum width=0.3\linewidth]
  (vfu) {VFU};
  \draw [thick, dotted] ($ (vwr.south west) + (2*0.5*\linewidth/8,0) $) -- (vfu.north west);
  \draw [thick, dotted] ($ (vwr.south west) + (3*0.5*\linewidth/8,0) $) -- (vfu.north east);

  \path (vfu) ++(4em,-4em) node [draw, thick, fill=green, rectangle, minimum height=2em, minimum width=4em] (r3) {R3};
  \path (vfu) ++(-4em,-4em) node [draw, thick, fill=green, rectangle, minimum height=2em, minimum width=4em] (r2) {R2};

  \path (vfu) ++(0,-7em) node [draw, thick, fill=light_blue, rectangle, minimum height=2em, minimum width=0.3\linewidth]
  (shuffler2) {Shuffler};

  \path (vfu.east) ++ (1em,0) node [anchor=west, trapezium, draw, thick, trapezium angle=60,minimum height=2em] (mux1) {};
  \path (vfu.west) ++ (-1em,0) node [anchor=east, trapezium, draw, thick, trapezium angle=60,minimum height=2em]
  (mux2) {};
  \path (vfu) ++(4em,3em) node [draw, thick, fill=sky_blue, rectangle, minimum height=2em, minimum width=4em] (r1) {R1};
  \path (vfu) ++(-4em,3em) node [draw, thick, trapezium, trapezium angle=-60, minimum height=2em] (mux3) {};
  \path (mux2) ++(0,3em) node [draw, thick, fill=sky_blue, rectangle, minimum height=2em, minimum width=4em] (r4) {R4};

  \draw[very thick] (mux2.north east |- mux2.south) coordinate (A) (r2.north) --++ (0,0.5em) -| (A);
  \draw[very thick] (r2.north) --++ (0,0.5em) -| (r3.north);
  \draw[very thick] (mux1.north west |- mux1.south) coordinate (B) (r3.north) --++ (0,0.5em) -| (B);
  \draw[very thick] (vfu.south) -- (shuffler2.north);
  \draw[very thick] (r2.south) -- (r2 |- shuffler2.north);
  \draw[very thick] (r3.south) -- (r3 |- shuffler2.north);
  \draw[very thick] (mux2.north west |- mux2.south) coordinate (C) (shuffler2.south) --++ (0,-0.5em) -| (C);
  \draw[very thick] (mux1.north east |- mux1.south) coordinate (D) (shuffler2.south) --++ (0,-0.5em) -| (D);
  \draw[very thick] (mux3.north east) -- (mux3.north east |- vwr.south);
  \draw[very thick] (mux3.south) -- (mux3 |- vfu.north);
  \draw[very thick] (r1.south) -- (r1 |- vfu.north);
  \draw[very thick] (r1.north) --(r1 |- vwr.south);
  \draw[very thick] (mux1.north) -- (mux1.north|-vwr.south east);
  \draw[very thick] (mux2.north) -- (r4.south);
  \draw[very thick] (r4.north) --++(0,0.5em) coordinate (E) -| (mux3.north west);


  \draw[decorate,decoration={brace,amplitude=10pt,raise=4pt},yshift=0pt]  (sram.north east) -- (vwr.south east)
  node [black,midway,xshift=1.5em,anchor=west] {4096 bits wide};

  \draw[decorate,decoration={brace,amplitude=10pt,raise=4pt},yshift=0pt]  (vwr.south east |- r1.north east) --
  (vwr.south east |- shuffler2.south east) node [black,midway,xshift=1.5em,anchor=west] {512 bits wide};
\end{tikzpicture}

%% file: parts/comparison.tex
\section{Conceptual comparison with other architectures}\label{sec:comparison} The architecture introduced here diverges
in several respects from two-dimensional systolic arrays and from one-dimensional GPU/vector arrays. In this section, we
focus on the conceptual distinctions. These distinctions are subsequently assessed in section~\ref{sec:experiments}.
Because Provet shares similarities with other data-parallel subword-oriented vector (1D) architectures, we will mainly
focus our detailed comparison there. In contrast, we will limit our comparison with systolic arrays to bandwidth and
scalability considerations.

Due to insufficient detailed microarchitectural data on commercial GPU internals, our comparison with GPUs will be
restricted to the memory hierarchy and the interconnect.

\subsection{Memory bandwidth and scalability}\label{sec:memory_bandwidth_comparison} Systolic arrays typically employ
SRAM widths considerably narrower than those proposed in this work. Common SRAM widths range from 128 to 512
\cite{Jouppi2017}. Section~\ref{sec:motivations} describes why wider SRAMs can be more energy-efficient; however,
another critical distinction lies in the memory bandwidth and how it scales.

\begin{figure}[tb]
    \begin{subfigure}{0.5\linewidth-1em}
        \begin{tikzpicture}
        \begin{axis}[
            ylabel={Bandwidth},
            xlabel={Number of PEs},
            grid=none,
            width=\textwidth,
            height=0.5\textwidth,
            xtick={0,20,...,100},
            tick pos=left,
            legend style={at={(1,0)}, anchor=south east, align=left, font=\footnotesize, },
            legend columns=3,
            ymode=log
        ]
            \addplot table [x=PEs, y=ProVeT, col sep=comma, mark=none] {data/bandwidth.csv};
            \addplot table [x=PEs, y=SA, col sep=comma, mark=none] {data/bandwidth.csv};
            \addplot [black] table [x=PEs, y=GPU, col sep=comma, mark=none] {data/bandwidth.csv};
            \legend{ProVeT, Systolic Array, GPU};
            \node at (axis cs:100,30) [anchor=center] {$\sim\beta N$};
            \node at (axis cs:5,20) [anchor=center] {\color{blue}$\sim\alpha N$};
            \node at (axis cs:100,5) [anchor=center] {\color{red}$\sim\sqrt N$};
        \end{axis}
        \end{tikzpicture}
        \caption{Available memory bandwidth scaling with increasing number of PEs ($N$). Normalized in terms of $N$. 
        Scaling factors shown near the curves, $\alpha > \beta$}
        \label{fig:bandwidth_comparison}
    \end{subfigure}%
    \hspace{2em}%
    \begin{subfigure}{0.5\linewidth-1em}
        \begin{tikzpicture}
        \begin{axis}[
            xlabel={Number of PEs}, 
            ylabel={Utilization},
            grid=none,
            width=\textwidth,
            height=0.5\textwidth,
            xtick={0,20,...,100},
            tick pos=left,
        ]
            \addplot table [x=size, y=ProVeT, col sep=comma, mark=none] {data/ideal_utilization.csv};
            \addplot table [x=size, y=SA, col sep=comma, mark=none] {data/ideal_utilization.csv};
            \addplot table [x=size, y=GPU2, col sep=comma, mark=none] {data/ideal_utilization.csv};
        \end{axis}
        \end{tikzpicture}
        \caption{Utilization scaling with increasing number of PEs for a 11x11 CONV kernel}
        \label{fig:pe_utilization}
    \end{subfigure}
    \begin{subfigure}{0.5\linewidth-1em}
        \begin{tikzpicture}
        \begin{axis}[
            xlabel={Number of PEs}, 
            ylabel={Interconnect},
            grid=none,
            width=\textwidth,
            height=0.5\textwidth,
            xtick={0,20,...,100},
            tick pos=left,
            legend style={at={(0,1)}, anchor=north west, align=left},
        ]
            \addplot table [x=size, y=3x3, col sep=comma, mark=none] {data/interconnect.csv};
            \addplot table [x=size, y=5x5, col sep=comma, mark=none] {data/interconnect.csv};
            \addplot table [x=size, y=7x7, col sep=comma, mark=none] {data/interconnect.csv};
            \addplot table [x=size, y=11x11, col sep=comma, mark=none] {data/interconnect.csv};
            \legend{3x3, 5x5, 11x11};
        \end{axis}
        \end{tikzpicture}
        \caption{Utilization of the local interconnect needed to fill the entirety of the systolic array for different
        kernel sizes}
        \label{fig:interconnect_utilization}
    \end{subfigure}%
    \caption{Comparison between Provet and an ideal systolic array}
    \label{fig:comparison}
\end{figure}
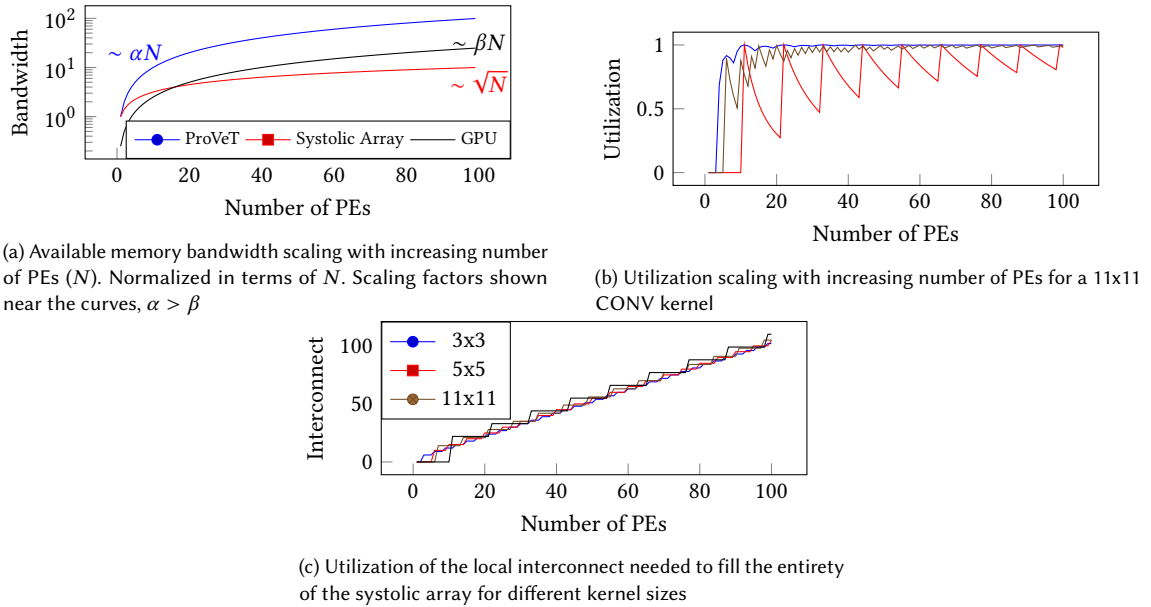

As discussed in section~\ref{sec:motivations}, the memory bandwidth of a systolic array is constrained by the square
root of the total number of PEs (Figure~\ref{fig:bandwidth_comparison}). This arises because the memory connects only to
the edge PEs, requiring data to be propagated through the interconnect network to reach those in the interior of the
array. Additionally, the utilization of these PEs depends on folding the input data across the array. As shown in
Fig.~\ref{fig:pe_utilization}, acceptable PE utilization is only reached when the array is large enough to accommodate
multiple folds of the input data. However, this conflicts with the earlier observation that memory bandwidth emerges as
a bottleneck when the array size increases.

In contrast, the proposed architecture ensures that memory bandwidth does not limit the utilization of the PEs (VFUs)
with a scaling number of PEs, as the memory bandwith increases accordingly with the number of PEs. Owing to the very
wide memory, Provet’s bandwidth scales linearly with the number of PEs, whereas in a systolic array it grows with the
square root of the PEs’ count (due to their square arrangement). Because of the asymmetry in the VWR ports, each VFU
needs $N$ cycles to consume the entire VWR, where $N$ is determined by the ratio of the SRAM’s width to the width of the
VFU. Instead, SAs rely on local data reuse to offset memory bandwidth limitations. When data reuse is limited, this
constrained bandwidth can lead to under-utilized PEs. This effect is particularly evident with depth-wise separable
convolutions, as will be shown in section~\ref{sec:experiments}.

\subsection{Interconnect organization}\label{sec:interconnect_comparison} A second major distinction from systolic
arrays lies in how Provet manages local data movement. Systolic arrays rely on nearest-neighbor interconnects, which
effectively route data in a linear sequence from one PE to the next. This approach is suitable when the dataflow can be
organized so that partial sums or other intermediate values move efficiently through adjacent PEs. However, when data
must traverse larger distances, the additional energy consumption and latency can become problematic. A typical scenario
arises when the input image dimensions do not align with the dimensions of the SA, requiring \emph{folding} across the
2D PE grid. In such cases, the once-linear dataflow is disrupted and data must jump across multiple PEs. The resulting
increase in communication latency can lower utilization when needed data arrives too late, forcing PEs to idle.

Provet, in contrast, uses shufflers to facilitate the necessary local data transfers to the VFUs. These shufflers can
link different VFU slots in a single cycle over significant distances. It is important to note that the maximum shuffler
range is determined at design time and should be guided by application profiling. If a large shuffling distance is
rarely used, the designer could restrict the maximum range and rely on two consecutive shuffle operations instead.
Long shuffling distances can lead to an increase in wire length and thus energy consumption. For this reason, the size
of the shuffler should be selecting based on a trade-off between the energy efficiency and the performance based on
profiling. As shown in Section~\ref{sec:mappings}, a suffle distance of 1 is sufficient to implemente the CONV and fully
connected kernels. It is important to note that the length of the wires will scale with the shuffle distance and not
with the size width of the shuffler (and Provet). If the shuffing distance stays the same, the length of the wires and
thus (normalized) energy consumption will not increase with an increase of width.

Moreover, local data movement in Provet generally proceeds vertically: data from a specific SRAM location is mostly
directed to the VFU slot that is vertically aligned, which reduces the frequency of horizontal transfers through the
shufflers. This characteristic is illustrated in the CONV example (section~\ref{sec:example}).

Figure~\ref{fig:interconnect_utilization} shows that, in an SA-based approach, interconnect usage scales with the size
of the array. Because data enters primarily at the array’s boundaries, more PEs entail longer paths through the network
before data reaches those located centrally. By contrast, in Provet, interconnect utilization remains relatively
constant since all PEs have direct access to memory through the VWR.

\subsection{Memory hierarchy}\label{sec:memory_hierarchy_comparison}
A primary distinction between Provet and other architectures lies in its memory hierarchy. Figure~\ref{fig:memory_hierarchy}
illustrates the various elements of that hierarchy as well as the interconnections among them.

\subsubsection{Systolic arrays}
They use a single global memory that supplies the entire array. They also incorporate input and output
registers for storing stationary data (input) and partial results (output) during computation. Data movement always
proceeds by (1) transferring data between the global memory and the input registers, (2) feeding data into the PE array,
(3) circulating data within the array for as long as reuse opportunities remain, and (4) eventually returning data from
the output registers to the global memory. As mentioned earlier, all mechanisms aimed at minimizing data access
fundamentally hinge on the presence of data reuse within the array, which depends on the specific application. Thus,
the memory hierarchy alone cannot ensure a fixed level of memory access reduction.

\subsubsection{ARA vector processor}
It's memory hierarchy is similar to that of Provet in that each PE is tied to a portion of the global memory, and a
register is placed between the global memory and the SIMD units. However, this register is a standard vector register
file rather than a VWR, and it primarily stores intermediate results between memory and SIMD operations. This difference
can hinder scalability, as a conventional (non-ultra-wide) register tends to become an energy bottleneck when the number
of PEs grows. The ARA interconnect also limits data movement among and within SIMD slices. While the general data
path—moving data from global memory to the register file and then to the SIMD units—is consistent with Provet, the
critical difference lies in the widths of each memory hierarchy element, which ultimately affects how effectively memory
accesses can be reduced.

\subsubsection{GPUs}
They feature a memory hierarchy that is similar to the one of systolic arrays. The main difference is that the memory
feeds into a 1D array of PEs, which means that the entire array can access the memory directly. GPUs do not feature any
of the intermediate elements that Provet and ARA provide, which means that the access to the main memory will not show
any reduction in terms of number of accesses. GPUs heavily rely on \emph{batching} to reduce the number of memory accesses
and to hide the latency of such accesses~\cite{delestrac2024multi}. This batching however introduces a significant increase in
the latency of the application. 
The data movement in GPUs is similar to the one in systolic arrays, where the data is (1) moved from the global memory
(through levels of cache) to the intermediate register file, then (2) to the SIMD unit, and finally (3) back to the
global memory. The only difference being the 1D organization of the PEs.
However, data movement between PEs has to go through caches. For example, with the Ampere architecture: PEs within the
same SM communicate through L1, and PEs from different SMs communicate through L2.
The nature of the register file in GPUs is completely different from the VWR in Provet. Regarding their sizes, the GPU
register file is closer to the global memory of Provet, while conceptually closer to the VWR.

The mismatches in both functionality (asymmetry) and scale shows that the VWR has no direct equivalent element in the
GPU memory hierarchy. Hence, this poses a challenge when comparing both architectures or when experimentally normalizing
the sizes of memory elements for comparison. In section~\ref{sec:experiments}, we provide different comparisons to
showcase that the improvements brought by Provet are valid in a wide range of scenarios.


%% file: parts/mappings.tex
\section{Mapping neural networks}\label{sec:mappings}
To analyze the capabilities of this architecture when executing common DNN algorithms, a set of DNN mappings has been
created and analyzed. We show how two of the most common types of layers can be mapped to the proposed architecture.
\begin{figure*}
\begin{subfigure}{0.61\linewidth}
    \centering
    \includegraphics[width=0.9\linewidth]{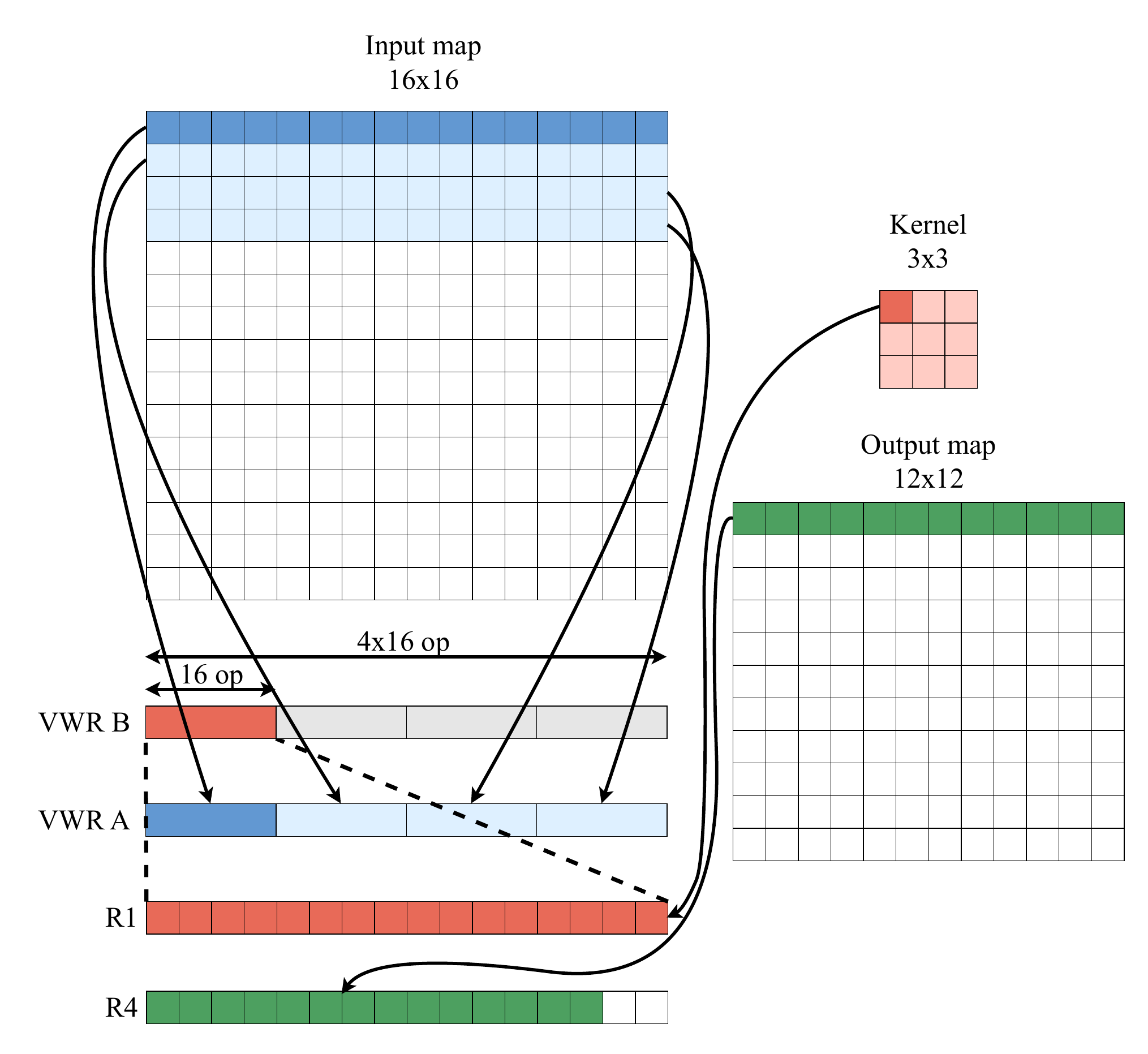}
    \caption{Data arrangement in VWRs and internal registers. The VWR A holds rows of the input map, 4 in this case. The
    VWR B holds the kernel. After the first row of the input map is consumed (5 iterations) the next one can be loaded
    (5th row)}
    \label{fig:data_placement}
\end{subfigure}%
\hfill
\begin{subfigure}{0.35\linewidth} 
    \centering
    \includegraphics[width=\linewidth]{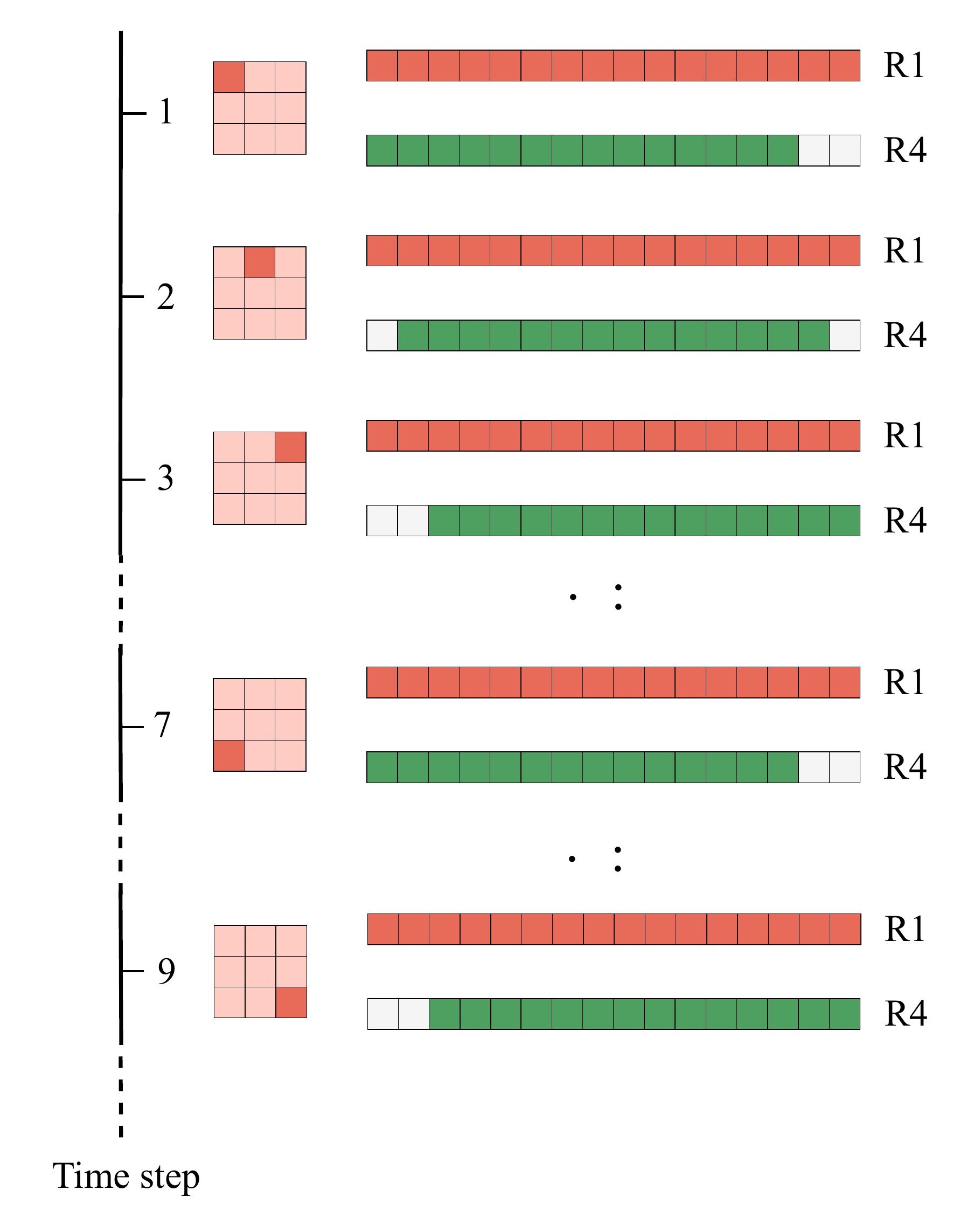}
    \caption{Data movement at different time steps. In the diagram we observe 25 iterations needed, one for each kernel
    pixel, to produce one row of the output. This operation is repeated 16 times, for a total of 400 iterations to
    generate the full output image}
    \label{fig:iterations}
\end{subfigure}
\caption{Example of a convolution layer}
\label{fig:example}
\end{figure*}
\subsection{Ilustrative example of a CONV layer}\label{sec:example}
In this section, we elaborate on a detailed example of a CONV layer mapping. The dimensions of both the CONV layer and
architecture have been chosen small to facilitate the illustrations and provide clarity to the reader on how the data
movement and organization are developed. The exact same principles that are presented in this example can be extended to
real-world dimensions.

We now proceed to map a 5x5 kernel and a 16x16 input map into an architecture where 16 operands per VFU, 64 operands
SRAM width, and a single VFU. Figure~\ref{fig:data_placement} shows how the data is arranged in the VWRs. VWR A holds
the first 4 rows of the input map and VWR B holds the entire kernel. 

The main idea of this dataflow is to hold the output map partial sum in R4 and iterate over all kernel pixels. In each
iteration, a kernel pixel is read and broadcasted to all positions of R1. Then R1 is multiplied by an entire row of the
image. Depending on which iteration, a different slice of the VWR will be selected. The result of the multiplication is
accumulated into R4. Finally, R4 contents are shifted one position to the right. This process is repeated five times
until the whole row of the kernel is processed. After that, R4 is shifted back to its original position, and the next
row of the kernel starts.

The following pseudo-code shows the data flow presented above using the descriptive instruction names. The descriptive
instruction names are used to facilitate the reading of the code and correspond to the instructions from
Table~\ref{tab:isa}.

\noindent\begin{minipage}{0.4\textwidth}
\begin{minted}{ruby}
GLV (i, i+2N+1) # load weights
for k in range(INPUT_HGHT-KERNEL_HEIGHT-1)
 for j in range(KERNEL_HEIGHT)
   for i in range(KERNEL_WIDTH)
    read_from_vwr(VWR=B, slice=j%5, mask=i)
    vfux_mult(in1=VWRA, in2=R1, out=R2)
    vfux_add(in1=R2, in2=R4, out=R4)
    shuffle(in=R4, out=R4, step=1)
   end
   shuffle(in=R4, out=R4, step=-4)
 end
end
\end{minted}
\label{fig:conv_code}
\end{minipage}

\subsection{Dealing with size mismatches}\label{sec:size-mismatch-mapping}
When the size of the hardware does not match the size of the image or kernel, the mappings have to be adjusted to deal
with the size mismatches. This process is typically called \emph{folding}. In this section we show how the mapping
concept presented in \ref{sec:example} can be adjusted to two different cases: image bigger than the hardware and image
smaller than the hardware. Considering that the proposed architecture targets large widths (e.g. 2046 bit) the first
situation will occur rarely but the second will very common.

\subsubsection{Image bigger than the width} 
\begin{figure}[tb]
    \centering
    \includegraphics[width=0.7\linewidth]{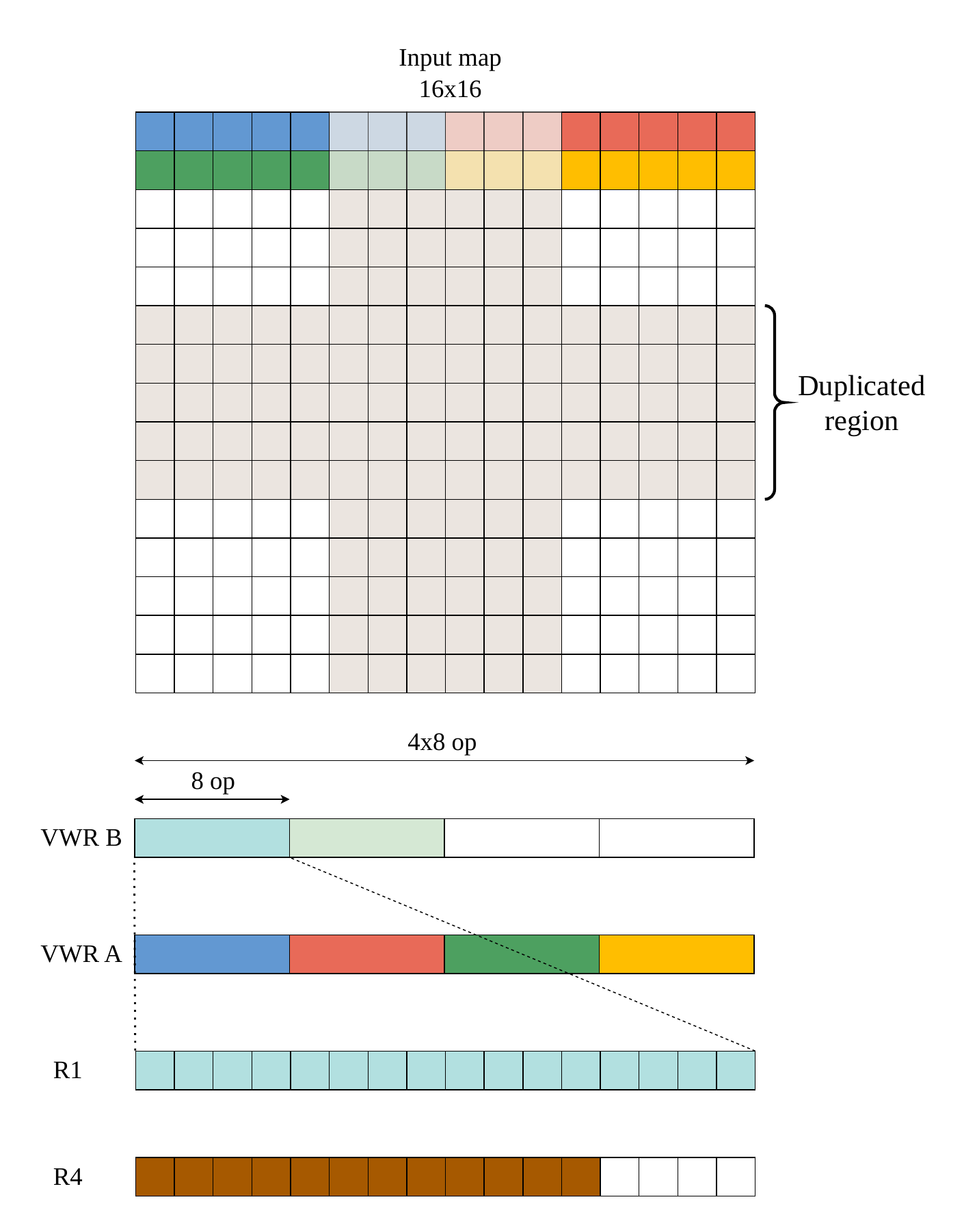}
    \caption{Fiting 1 input map wider than the VFU}
    \label{fig:overlaping}
\end{figure}

the solution for this situation is to partition the image into smaller
parts. As shown in Figure~\ref{fig:overlaping}, the input image can be partitioned into 4 smaller sub-images. Those
sub-images can then be processed independently as a normal (smaller) image. The downside of this solution is that a
portion of the image, marked in gray, will have to be duplicated. This is caused by the sliding nature of the
convolution kernel. The size duplicated area is given by the size of the kernel. It is important to note that the
overhead caused by this duplication will be practically negligible. As previously mentioned, the typicall widths found
in this architecture will be 1024-2046bit or higher. Considering that the bigger kernels that are commonly used in CNNs
are 11x11, the duplicated portion of the image will be less than 5\% of the total image size.

\subsubsection{Image smaller than the width} 
\begin{figure}[tb]
    \centering
    \includegraphics[width=0.7\linewidth]{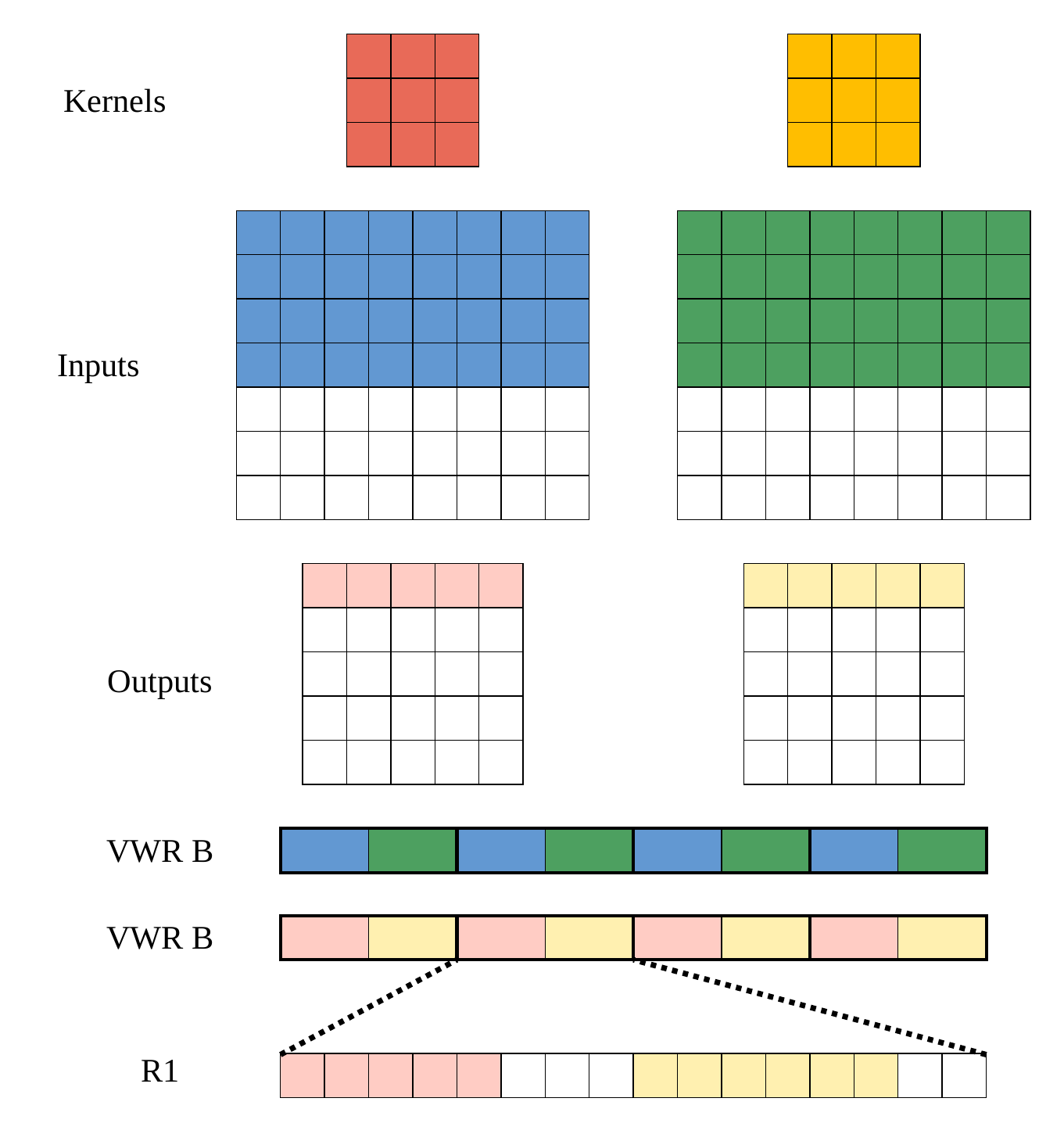}
    \caption{Fitting 2 kernels in 1 VFU}
    \label{fig:multiple_kernels}
\end{figure}

the solution for this situation is to fit two separate images into the VWR
and VFUs simultaneously. Figure~\ref{fig:multiple_kernels} shows how two images and kernels are merged into the VWRs and
VFUs. This particular example shows how the two images, when combined, fit exactly into the width of the VWRs. This is,
however not the general case and some empty portions will often be left unfilled. These empty slots will lead to an
under-utilization in the VFUs. However, it is easy to see how the utilization will asymptotically reach 100\% for large
widths.

\subsection{Template system}\label{sec:template}
Using the bare instructions to write complex DNN applications can be very complex and slow since it would require
writing the equivalent of assembly code. We propose a template system with a basic set of templates that implement basic
functions commonly used in DNNs, such as multi-dimensional convolutions, max pool, fully connected, etc. This
\emph{macros} are provided as a library that can later be used to construct more complex algorithms. The templates
incorporate the two aspects of a mapping: the instructions and the memory layout. These templates aim to provide a
similar interface layer that libraries such as PyTorch or TensorFlow provide for GPU architectures. 

%% file: parts/results.tex
\section{Experiments and results}\label{sec:experiments}
To validate the claims put forward for the proposed architecture, we compare several mappings against two systolic
array (SA) architectures—Eyeriss and TPU—alongside a vector processor (ARA) and an Nvidia Ampere GPU.

For TPU and Eyeriss, mappings are generated using the ZigZag design space exploration framework, along with the hardware
and mapping templates provided for those architectures to align with their respective publications \cite{Jouppi2017,
10.1109/ISCA.2016.40}. The ARA mappings are derived from the code snippets provided by the original authors. GPU
mappings are implemented using TensorFlow and the cuDNN library, the metrics are extracted using the methodology
proposed by Delestrac et.al.~\cite{delestrac2024analyzing}. The Provet mappings are produced according to the
methodology introduced in section~\ref{sec:mappings}.

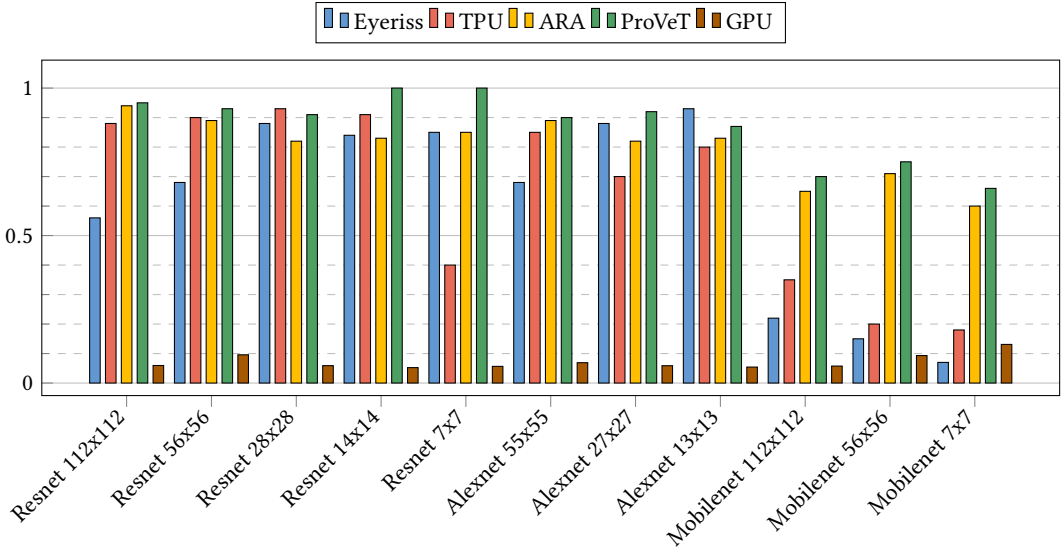
\begin{figure}[tb]
  \centering
  \begin{tikzpicture}
    \begin{axis}[
      ybar,
      width=\linewidth,
      height=0.4\linewidth,
      enlargelimits=0.1,
      legend style={at={(0.5,1.05)}, anchor=south, legend columns=-1},
      symbolic x coords={Resnet 112x112, Resnet 56x56, Resnet 28x28, Resnet 14x14, Resnet 7x7, Alexnet 55x55,
      Alexnet 27x27, Alexnet 13x13, Mobilenet 112x112, Mobilenet 56x56, Mobilenet 7x7},
      xtick=data,
      xticklabel style={rotate=45, anchor=east},
      minor y tick num=4,
      tick pos=left,
      bar width=4pt,
      ymajorgrids=true,
      yminorgrids=true,
      minor grid style=dashed,
      ]
      \addplot [draw, fill=sky_blue]    table[x=Name, y=Eyeriss, col sep=comma] {data/pe_utilization.csv};
      \addplot [draw, fill=brick]       table[x=Name, y=TPU, col sep=comma] {data/pe_utilization.csv};
      \addplot [draw, fill=yellow]      table[x=Name, y=ARA, col sep=comma] {data/pe_utilization.csv};
      \addplot [draw, fill=green]       table[x=Name, y=ProVeT, col sep=comma] {data/pe_utilization.csv};
      \addplot [draw, fill=dark_yellow] table[x=Name, y=GPU, col sep=comma] {data/pe_utilization.csv};
      \legend{Eyeriss,TPU,ARA,ProVeT,GPU}
    \end{axis}
  \end{tikzpicture}
  \caption{PE utilization for different layers of state-of-the-art CNNs.}
  \label{fig:cnn_results}
\end{figure}

Figure~\ref{fig:cnn_results} presents the utilization results for various layers of ResNet, AlexNet, and MobileNet.
Provet’s utilization figures derive from manual mappings, whereas for the ZigZag-generated mappings, utilization is
computed as the ratio of the theoretical minimum latency to the observed latency, as given in
equation~\eqref{eq:utilization_zigzag}. In this equation, $L_{\text{min}}$ is the minimum latency, and $L_{\text{real}}$
is the actual latency of the mapping. The minimum latency is obtained by summing all MAC operations within a layer and
dividing by the total number of PEs. The real workload-dependent latency is determined by analyzing the DSE results to
count the number of cycles used to process the layer. The GPU utilization is calculated by considering only the
memory-related stalls, as the other architectures do not include the control overhead in their models. This is done by
collecting the amount of control stalls and their relative occurance, which is found to be \SI{75.6}{\percent} of the
total stalls (see Fig.~\ref{fig:a100_control_stalls}). The GPU utilization is then scaled up by this factor to
provide a fair comparison with the other architectures.


The compute-to-memory instruction ratio is computed by recording the total number of compute instructions (executed in
the VFU) and the total number of memory instructions (executed in the global data buffer), and then forming the quotient.
Equation~\eqref{eq:compute_ratio} captures this formally. A high ratio indicates that the architecture performs more
computations per memory access, while a low ratio implies that the architecture is limited by memory bandwidth, as the
PEs cannot remain fully occupied.

\bigskip
\noindent\begin{minipage}{0.5\textwidth}
    \begin{equation}
      \label{eq:utilization_zigzag}
      U = \frac{L_{\text{min}}}{L_{\text{real}}}
    \end{equation}
\end{minipage}%
\begin{minipage}{0.5\textwidth}
    \begin{equation}
      \label{eq:compute_ratio}
      CMR = \frac{N_{\text{compute}}}{N_{\text{memory}}}
    \end{equation}
\end{minipage}

\medskip
To summarize, Table~\ref{tab:improvement} reports Provet’s improvements relative to other architectures for each tested
configuration. These improvements are calculated as the ratio between Provet’s average utilization and the best
utilization achieved by the other architectures, with the same procedure used to determine the compute-to-memory ratio.

\begin{table}[t]
    \centering
    \caption{Improvement of Provet over other architectures for ResNet (RN), AlexNet (AN) and MobileNet (MN) convolution
    layers for different kernel sizes. \textcolor{red}{Red values are preliminary estimations}}
    \label{tab:improvement}
    \begin{tabular}{lllllllll}
    \toprule
    \multirow{2}{*}{Layer} & \multicolumn{4}{c}{\textbf{Utilization improvement}} & \multicolumn{4}{c}{\textbf{CMR improvement}} \\
                        & \textbf{Eyeriss}  & \textbf{TPU}  & \textbf{ARA}  & \textbf{GPU} & \textbf{Eyeriss}      & \textbf{TPU}       & \textbf{ARA}      & \textbf{GPU} \\ \midrule
     RN\_112x112        & x1.70           & x1.08        & x1.01        & x15.97       & x4.09           & x3.00        & x1.36        & x1.25       \\
     RN\_56x56          & x1.37           & x1.03        & x1.04        & x9.71        & x3.63           & x3.00        & x1.24        & x1.21       \\
     RN\_28x28          & x1.03           & x0.98        & x1.11        & x15.42       & x4.14           & x3.03        & x1.28        & x1.11       \\
     RN\_14x14          & x1.19           & x1.10        & x1.20        & x19.12       & x4.00           & x3.29        & x1.31        & x1.26       \\
     RN\_7x7            & x1.18           & x2.50        & x1.18        & x17.67       & x3.60           & x3.33        & x1.53        & x1.61       \\
     \midrule
     AN\_55x55          & x1.32           & x1.06        & x1.01        & x13.04       & x3.95           & x3.48        & x1.50        & x1.16       \\
     AN\_27x27          & x1.05           & x1.31        & x1.12        & x15.65       & x4.24           & x3.07        & x1.41        & x1.20       \\
     AN\_13x13          & x0.94           & x1.09        & x1.05        & x16.05       & x4.09           & x3.00        & x1.48        & x1.00       \\
     \midrule
     MN\_112x112        & x3.18           & x2.00        & x1.08        & x12.15       & x25.00          & x15.00       & x3.13        & x2.14       \\
     MN\_56x56          & x5.00           & x3.75        & x1.06        & x8.05        & x19.50          & x15.60       & x2.69        & x3.00       \\
     MN\_7x7            & x9.43           & x3.67        & x1.10        & x5.04        & x24.67          & x18.50       & x2.96        & x3.08       \\
     \bottomrule
    \end{tabular}
\end{table}

All the discussed architectures, including the proposed one, are capable of attaining considerable levels of utilization
during the execution of conventional convolution operations in popular neural networks like AlexNet and ResNet. However,
it can be observed that SAs utilization can dip below 75\% when it comes to larger kernels.\bigskip

This drop in utilization can be attributed to two main factors: the rigid interconnect and the size mismatches. Refeer
to section~\ref{sec:size-mismatch} for a more detailed explanation.

%

A more significant difference in utilization can be observed in the MobileNet layers which exhibits a less regular
data-flow pattern with less exploitable data reuse thus imposing greater demands on memory bandwidth. When the memory
bandwidth is not sufficient to keep all PEs the utilization degrades. This situation can be observed for all MobileNet
layers in Figure~\ref{fig:cnn_results}. Eyeriss and TPU utilization collapse to very low numbers due to a severe memory
bottleneck. However, Provet and ARA can maintain reasonable utilization numbers thanks to the much higher memory
bandwidth available.\bigskip

\begin{figure}[tb]
  \centering
  \begin{tikzpicture}
    \begin{axis}[
      ybar,
      width=\linewidth,
      height=0.4\linewidth,
      enlargelimits=0.1,
      legend style={at={(0.5,1.05)}, anchor=south, legend columns=-1},
      symbolic x coords={Resnet 112x112, Resnet 56x56, Resnet 28x28, Resnet 14x14, Resnet 7x7, Alexnet 55x55,
      Alexnet 27x27, Alexnet 13x13, Mobilenet 112x112, Mobilenet 56x56, Mobilenet 7x7},
      xtick=data,
      xticklabel style={rotate=45, anchor=east},
      tick pos=left,
      bar width=4pt,
      ymode=log,
      ymajorgrids=true,
      yminorgrids=true,
      minor grid style=dashed,
      ]
      \addplot [draw, fill=sky_blue]    table[x=Name, y=Eyeriss, col sep=comma] {data/compute_ratio.csv};
      \addplot [draw, fill=brick]       table[x=Name, y=TPU, col sep=comma] {data/compute_ratio.csv};
      \addplot [draw, fill=yellow]      table[x=Name, y=ARA, col sep=comma] {data/compute_ratio.csv};
      \addplot [draw, fill=green]       table[x=Name, y=ProVeT, col sep=comma] {data/compute_ratio.csv};
      \addplot [draw, fill=dark_yellow] table[x=Name, y=GPU, col sep=comma] {data/compute_ratio.csv};
      \legend{Eyeriss,TPU,ARA,ProVeT,GPU}
    \end{axis}
  \end{tikzpicture}
  \caption{Compute-to-memory access ratio for different layers of state-of-the-art CNNs.}
  \label{fig:compute-ratio}
\end{figure}
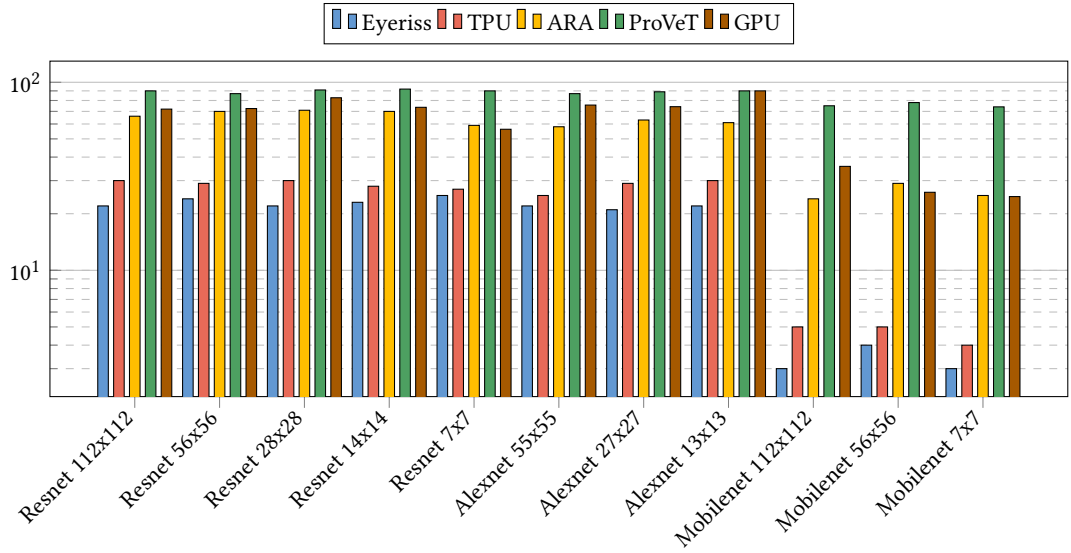

In addition to the PE utilization shown in Figure~\ref{fig:cnn_results}, the compute-to-memory access ratio was also
evaluated for the same mappings. These findings are presented in Figure~\ref{fig:compute-ratio}. Notably, Provet
exhibits a significantly higher compute-to-memory ratio, indicating that more computations are performed for each global
memory access compared to Eyeriss, TPU, and ARA.

This distinction is especially pronounced for MobileNet mappings, where the level of data reuse is substantially lower
than that in ResNet or AlexNet. As described in section \ref{sec:memory_assymetry}, SAs depend on data reuse as the
primary method to minimize memory access, thereby maximizing the compute-memory ratio. However, Provet utilizes the
memory asymmetry of the VWR to minimize memory access operations. Although Provet also benefits from local data reuse,
its architectural differences allow for a better compute-to-memory ratio even under conditions of reduced data reuse.

The overall findings match the expectations derived from the theoretical analysis of the architectures presented in the
earlier sections. The mixture of a 1D organization, the intermediate memory (VWR) and the data shufflers, allow Provet
to achieve higher utilization and better compute-to-memory ratios than the other architectures.
\begin{table}[tb]
  \caption{Memory accesses and latency for different architectures. All values are scaled \cite{Sarangi2021} to an equivalent 28 nm technology node and 200 MHz clock frequency.}
  \resizebox{\textwidth}{!}{%
    \begin{tabular}{@{}llllllllllll@{}}
      \toprule
      \multirow{2}{*}{\textbf{Layer}} & \multirow{2}{*}{\textbf{MOPS}} & \multicolumn{2}{c}{\textbf{Eyeriss}} & \multicolumn{2}{c}{\textbf{TPU}}  & \multicolumn{2}{c}{\textbf{ARA}}  & \multicolumn{2}{c}{\textbf{A100}} & \multicolumn{2}{c}{\textbf{Provet}} \\ \cmidrule(l){3-12} 
                                      &                                & \textbf{Reads}   & \textbf{Latency}  & \textbf{Reads} & \textbf{Latency} & \textbf{Reads} & \textbf{Latency} & \textbf{Reads} & \textbf{Latency} & \textbf{Reads}  & \textbf{Latency}  \\ \midrule
      Resnet 112x112                  & 236.0                          & 22.434           & 9.231             & 33.891         & 0.320            & 15.125         & 5.657            & 90.287         & 1.757            & 6.611           & 0.193             \\
      Resnet 56x56                    & 231.2                          & 22.093           & 9.035             & 33.058         & 0.315            & 14.820         & 5.516            & 88.416         & 1.713            & 6.454           & 0.189             \\
      Resnet 28x28                    & 115.6                          & 11.025           & 4.492             & 16.587         & 0.156            & 7.398          & 2.777            & 44.302         & 0.856            & 3.223           & 0.095             \\
      Resnet 14x14                    & 115.6                          & 11.072           & 4.536             & 16.493         & 0.157            & 7.414          & 2.785            & 44.258         & 0.861            & 3.222           & 0.095             \\
      Resnet 7x7                      & 115.6                          & 11.067           & 4.551             & 16.609         & 0.157            & 7.344          & 2.752            & 44.230         & 0.859            & 3.189           & 0.095             \\
      \midrule
      Alexnet 55x55                   & 210.8                          & 20.156           & 8.257             & 30.189         & 0.286            & 13.456         & 5.029            & 80.055         & 1.550            & 5.834           & 0.171             \\
      Alexnet 27x27                   & 895.8                          & 85.803           & 34.885            & 127.607        & 1.223            & 57.337         & 21.333           & 342.714        & 6.639            & 24.942          & 0.729             \\
      Alexnet 13x13                   & 299.0                          & 28.512           & 11.630            & 42.560         & 0.406            & 19.174         & 7.107            & 114.604        & 2.211            & 8.363           & 0.244             \\
      \midrule
      Mobilenet 112x112               & 0.7                            & 0.131            & 1.125             & 0.191          & 0.435            & 0.088          & 0.954            & 0.512          & 3.059            & 0.038           & 0.339             \\
      Mobilenet 56x56                 & 1.8                            & 0.340            & 0.768             & 0.515          & 0.510            & 0.231          & 1.071            & 1.374          & 3.651            & 0.101           & 0.403             \\
      Mobilenet 7x7                   & 0.5                            & 0.090            & 0.689             & 0.131          & 0.218            & 0.057          & 0.887            & 0.343          & 2.089            & 0.025           & 0.230             \\ \bottomrule
    \end{tabular}
  }
  \label{tab:memory_access_latency}
\end{table}

\begin{figure}[tb]
  \centering
  \begin{subfigure}{0.55\linewidth}
  \begin{tikzpicture}
    \begin{axis}[
      ybar,
      width=\linewidth,
      height=0.6\linewidth,
      enlargelimits=0.1,
      legend style={at={(1,1)}, anchor=north east, legend columns=1, fill=none, draw=none},
      symbolic x coords={Eyeriss, TPU, ARA, A100, ProVeT},
      xtick=data,
      xticklabel style={rotate=45, anchor=east},
      tick pos=left,
      bar width=4pt,
      ymode=log,
      ymajorgrids=true,
      yminorgrids=true,
      minor grid style=dashed,
      ]

      \addplot [draw, fill=sky_blue] coordinates {(Eyeriss, 46.73) (TPU, 69.64) (ARA, 31.27) (A100, 186.83) (ProVeT, 13.61)};
      \addplot [draw, fill=brick]    coordinates {(Eyeriss, 19034.98) (TPU, 666.30) (ARA, 11639.25) (A100, 3623.20) (ProVeT, 398.47)};
      \legend{Reads (MB), Latency (s)}
    \end{axis}
  \end{tikzpicture}
  \caption{Average memory accesses and latency for different architectures, weighted
  by the number of operations from Table~\ref{tab:memory_access_latency}.}
  \label{fig:memory_access}
  \end{subfigure}%
  \hfill
  \begin{subfigure}{0.4\linewidth}
    \centering
    \begin{tikzpicture}
      \pie[
      radius=1.6,
      sum=auto,
      text=legend,
      color={sky_blue, brick, green, yellow, turquoise, gray}
      ]{
        75.64/Memory,
        15.73/Control,
        8.62/Others
      }
  \end{tikzpicture}
  \vspace{3em}
  \caption{Distribution of control stalls for the A100 GPU.}
  \label{fig:a100_control_stalls}
  \end{subfigure}
\caption{}
\end{figure}

In addition to the utilization and compute-to-memory ratio, which are the primary metrics for evaluating the
improvements of Provet innovations, we have also considered the memory accesses and latency for the same convolutional
kernels. These metrics are presented in Table~\ref{tab:memory_access_latency} and summarized in
Figure~\ref{fig:memory_access}. The memory accesses are calculated as the number of reads from the global memory, as
defined in \ref{fig:smemory_hierarchies}. The latency is calculated as the time to process an individual layer,
including the time to read the input data, perform the computations, and write the results back.

The results show a significant reduction in global memory accesses for Proved when compared to other architectures, more 
notably for systolic arrays and GPUs as anticipated. The latency\footnote{Latency results are scaled to a 200 MHz clock
and equivalent 28 nm technology node.} is also significantly lower for Provet, which is a direct consequence of the high
utilization achieved in its processing elements. An important observation is that vector processors (Provet and ARA) have
the lowest total amount of memory accesses, which confirms the motivations presented in section~\ref{sec:motivations}.
Finally, GPUs provide a very low latency, but fail to achieve low memory accesses when batch size is set to 1, as its
ability to exploit data reuse is severly hindered. Other architectures on the other hand, can exploit some level of data
reuse even when the batch size is set to 1.

%% file: parts/conclusions.tex
\section{Conclusions} This paper has proposed an innovative memory hierarchy that effectively addresses the challenge of
memory bandwidth in data-parallel AI/ML applications. By introducing three levels of on-chip memory hierarchy - local,
intermediate, and global - with ultra-wide registers and specialized data-shufflers, the proposed architecture improves
versatility and adaptivity to varying data-parallel applications, even for those with limited data reuse. The mapping of
a representative data parallel application, such as a convolutional neural network, to the proposed architecture
highlights its superiority over existing systolic array-based ML/AI architectures in terms of memory bandwidth
bottlenecks. We demonstrate how a set of mappings can be used to construct a library that enables developers to quickly
and easily map their application onto the proposed wide-shallow memory hierarchy. By leveraging this library, developers
can accelerate the development process and focus on optimizing their applications for performance and power rather than
spending significant time on memory hierarchy design. This approach not only simplifies the design process, but also
enables the development of highly efficient AI/ML applications that take full advantage of the proposed memory
hierarchy. Additionally, our study showed that scaling a wide-shallow memory is a better approach than using a square
memory design.